\documentclass{aa}
\usepackage{xcolor}
\usepackage[breaklinks=true, colorlinks, citecolor={blue}, urlcolor={blue}]{hyperref}
\usepackage{graphicx}
\graphicspath{ {./images/} }
\usepackage{amssymb}
\usepackage{booktabs} 
\usepackage{afterpage}
\usepackage[varg]{txfonts}
\usepackage{supertabular}
\usepackage[normalem]{ulem}
\usepackage{csquotes}
\usepackage{mathtools}
\usepackage{float}
\usepackage{natbib}
\usepackage{footmisc}
\usepackage{soul}

\usepackage[switch]{lineno}
\usepackage{multirow}
\usepackage{multicol}
\usepackage{amsfonts}
\usepackage{eucal}
\usepackage{tabularx}
\usepackage{siunitx}
\usepackage{geometry}
\usepackage{pdflscape}
\usepackage{subcaption} 
\usepackage{comment}
\usepackage{pifont}
\usepackage{tikz}
\newcommand{\redbox}{%
  \tikz\fill[red] (0,0) rectangle (1cm,0.3cm);%
}
\newcommand{\greenbox}{%
  \tikz\fill[green] (0,0) rectangle (1cm,0.3cm);%
}

\usepackage{appendix}
\usepackage{natbib,twoopt}
\bibpunct{(}{)}{;}{a}{}{,} 
\makeatletter
\newcommandtwoopt{\citeads}[3][][]{\href{http://adsabs.harvard.edu/abs/#3}%
{\def\hyper@linkstart##1##2{}%
\let\hyper@linkend\@empty\citealp[#1][#2]{#3}}} 
\newcommandtwoopt{\citepads}[3][][]{\href{http://adsabs.harvard.edu/abs/#3}%
{\def\hyper@linkstart##1##2{}%
\let\hyper@linkend\@empty\citep[#1][#2]{#3}}}
\newcommandtwoopt{\citetads}[3][][]{\href{http://adsabs.harvard.edu/abs/#3}%
{\def\hyper@linkstart##1##2{}%
\let\hyper@linkend\@empty\citet[#1][#2]{#3}}}
\newcommandtwoopt{\citeyearads}[3][][]%
{\href{http://adsabs.harvard.edu/abs/#3}
{\def\hyper@linkstart##1##2{}%
\let\hyper@linkend\@empty\citeyear[#1][#2]{#3}}}

\makeatother

\bibpunct{(}{)}{;}{a}{}{,}

\bibpunct{(}{)}{;}{a}{}{,} 

\newcommand{\order}[1]{} 

\def\gsim{ \lower .75ex \hbox{$\sim$} \llap{\raise .27ex \hbox{$>$}} }

\def\lsim{ \lower .75ex\hbox{$\sim$} \llap{\raise .27ex \hbox{$<$}} }

\begin{document}
\title{Detection of TeV emission during early afterglow from poorly localized GRBs with ground based IACTs}
\author{S. Macera\inst{\ref{inst1},\ref{inst2},\ref{inst15} }\thanks{\email{\href{mailto:samanta.macera@gssi.it}{samanta.macera@gssi.it}}} 
\and B. Banerjee\inst{\ref{inst1},\ref{inst2}, \ref{inst3}}\thanks{\email{\href{mailto:biswajit.banerjee@gssi.it}{biswajit.banerjee@gssi.it}}}
\and J. Green \inst{\ref{inst5},\ref{inst6}}
\and M. Seglar-Arroyo\inst{\ref{inst4}}
\and G. Oganesyan \inst{\ref{inst1},\ref{inst2}, \ref{inst3}}
\and P. Tiwari\inst{\ref{inst1},\ref{inst2}}
\and A. Ierardi\inst{\ref{inst1},\ref{inst2}}
\and M. Branchesi\inst{\ref{inst1},\ref{inst2}, \ref{inst3}}
\and F. Aharonian \inst{\ref{inst7},\ref{inst8}, \ref{inst9}}
\and S. Mohnani  \inst{\ref{inst10}}
\and D. Miceli \inst{\ref{inst1}, \ref{inst11}, \ref{inst12}}
\and F. Sch\"ussler\inst{\ref{inst13}} 
\and A. Berti \inst{\ref{inst14}}}
\institute{Gran Sasso Science Institute, Viale F. Crispi 7, I-67100, L'Aquila (AQ), Italy
\label{inst1}
\and
INFN - Laboratori Nazionali del Gran Sasso, I-67100, L’Aquila (AQ), Italy\label{inst2}
\and
INAF - Osservatorio Astronomico d’Abruzzo, Via M. Maggini snc, I-64100 Teramo, Italy \label{inst3}
\and
Institut de Fisica d'Altes Energies (IFAE), The Barcelona Institute of Science and Technology, Campus UAB, 08193 Bellaterra (Barcelona), Spain \label{inst4}
\and
Max-Planck-Institut f{\"u}r Physik, Boltzmannstr. 8, 85748 Garching, Germany \label{inst5}
\and
Technische Universit{\"a}t M{\"u}nchen, 85748 Garching, Germany \label{inst6}
\and 
Dublin Institute for Advanced Studies, 31 Fitzwilliam Place, Dublin 2, Ireland \label{inst7}
\and 
Max-Planck-Institut f\"ur Kernphysik, Saupfercheckweg 1, 69117 Heidelberg, Germany \label{inst8}
\and 
Yerevan State University, 1 Alek Manukyan St, Yerevan 0025, Armenia \label{inst9}
\and
Department of Astronomy, Astrophysics and Space Engineering Indian Institute of Technology Indore, Simrol, Khandwa Road, Indore 453552, Madhya Pradesh, India. \label{inst10} 
\and 
INFN - Sezione di Padova I-35131 Padova Italy
\label{inst11}
\and
INAF - Osservatorio Astronomico di Brera, Via Emilio Bianchi 46, 23807 Merate, Italy
\label{inst12}
\and
IRFU, CEA, Universit\'e Paris-Saclay, F-91191 Gif-sur-Yvette, France\label{inst13}
\and 
Max-Planck Institute for Physics,
Boltzmannstrasse 8, 85748 Garching, Germany\label{inst14}
\and
INAF - Osseravtorio Astronomico di Roma, Via Frascati 33, I-00078 Monte Porzio Catone, Italy \label{inst15}
}

\authorrunning{Macera et. al 2026}

\abstract{
Gamma-ray bursts (GRBs) are among the most luminous and rapidly evolving transients in the Universe. While space-based instruments have extended GRB observations up to energies of $\sim$100 GeV, the detection of very-high-energy (VHE; $E>100$ GeV) emission from ground-based telescopes, especially during prompt or/and the early afterglow phase, remains challenging. These difficulties arise from the rapid temporal decay of GRB afterglows, strong attenuation by the extragalactic background light (EBL), observational latency, and the typical poor sky localization provided by MeV-detectors such as Fermi/GBM. In this work, we investigate the prospects for detecting TeV ($\sim$100 GeV--1 TeV) emission from poorly localized GRBs by adopting optimized follow-up strategies based on rapid tiling of large localization regions. We simulate a realistic population of GRBs informed by more than fifteen years of Fermi/GBM and Swift/XRT detections and recent progresses in the afterglow emission modeling. Using these simulations, we evaluate the detectability of GRB early afterglows by the next-generation Imaging Atmospheric Cherenkov Telescopes, equipped with larger field-of-view (FoV), as a function of latency, exposure time, and observational strategy. Our strategy can significantly enhance the detection rate; for instruments such as ASTRI and LACT, it increases by up to a factor of two compared to strategies limited to well-localized (Swift-like) events. For CTAO, our proposed approach provides up to four VHE detections per year.}


\maketitle

\section{Introduction}\label{sec:intro}\par

Gamma-ray bursts (GRBs) are among the most luminous and rapidly evolving transients in the Universe. They appear as intense flashes of keV-MeV gamma-rays lasting from a fraction of a second to several minutes, commonly interpreted as the prompt emission produced within ultra-relativistic jets launched either after the collapse of massive stars or following binary neutron star mergers. After this brief prompt emission phase, the outflow decelerates as it sweeps up the surrounding medium and gives rise to a longer-lived multi-wavelength afterglow. The interaction between the jet and the environment results in a relativistic forward shock that accelerates electrons, which then radiate across the electromagnetic spectrum through the synchrotron and Inverse-Compton (IC) processes.
\par
The very first detection of high-energy (HE) radiation ($>$100 MeV) associated with GRBs have been reported by EGRET onboard the  \textit{ Compton Gamma Ray Observatory} \citep{Fishman1995,Kaneko2008}. Over the last 18 years, observations by {the} Large Area Telescope (LAT) onboard the Fermi Gamma-ray Space Telescope have revealed that some GRBs exhibit a distinct HE spectral component that extends into the GeV domain during or shortly after the prompt emission phase \citep[e.g.]{Abdo2009a,Abdo2009c,Ackerman2013,Tang2015,Yassine2017,Vianello2018,2019ApJ...878...52A,Chand2020,Mei2022,Ravasio2024,Macera:2025wrv}. These detections suggest the presence of a second radiative component-potentially linked to the onset of the external shock-that overlaps with the classical prompt emission \citep{Zou2009,Gao2009,Kumar2009,Ghisellini2010,Ghirlanda2010}. However, in the absence of observations at {very-high-energies} (VHE; E$>$100\,GeV), the nature, temporal evolution, and energetics of this component remain poorly constrained. In particular, the transition from prompt to afterglow emission, the relative contributions of synchrotron and IC processes \citep{Sari:2000zp}, and the expected luminosity and duration of the VHE radiation in the earliest phases remain open questions.\par
A major advance came with the first detections of TeV emission from GRBs by ground-based Imaging Atmospheric Cherenkov Telescopes (IACTs). MAGIC and H.E.S.S. independently discovered VHE radiation from GRB 190114C and GRB 180720B, respectively \citep{Abdalla:2019dlr,MAGIC:2019irs,MAGIC:2019lau}, revealing that GRB afterglows can produce photons well above 100 GeV on timescales from minutes to hours after the trigger. TeV emission was also observed from GRB 190829A \citep{2021Sci...372.1081H} and GRB 201216C \citep{Abe:2023nhj}. For almost all of these events, there was a simultaneous detection of VHE and X-ray data. 
However, despite these remarkable TeV observations, multiwavelength spectra remained poorly constrained, given the absence of data in the MeV-GeV range. The situation changed with the breakthrough detection of the exceptionally bright "BOAT" GRB (Brightest of all time), GRB~221009A, for which LHAASO observed TeV photons temporarily coincident with the prompt emission phase through the onset of the afterglow \citep{LHAASO:2023kyg,LHAASA:2023pay}. Data were available for the first time across the full keV-TeV energy range \citep{Williams:2023sfk,2023ApJ...952L..42L,Tavani:2023hup,Fermi-LAT:2024jox,CTAO-LST:2025wee}. A distinct second spectral component was clearly identified and its peak energy was determined \citep{Banerjee:2024hxp}. This continuous broadband coverage was crucial for studying the prompt to afterglow evolution and for constraining the microphysical parameters of the emitting region \citep[see also][]{Derishev:2023xyx,Khangulyan:2023srq,Foffano:2024hnp,Barnard:2025zjx}. Early time VHE observations are still limited by large localization uncertainties from triggering instruments, General Coordinates Network circulars (GCN) delays, repointing of IACTs, and huge attenuation by extragalactic background light (EBL). As demonstrated by \cite{Mohnani:2025qfg}, GRB 230812B illustrates how the absence of accurate localization can prevent VHE follow-up observations, even when the burst sky region is visible to the telescope and the emisison is detectable.

\par

From an observational perspective, GRB triggering instruments can be broadly divided into two main categories. The first class consists of instruments providing precise sky localization, typically at the level of a few arcminutes. This category includes instruments such as the Burst Alert Telescope (BAT) onboard the Neil Gehrels Swift Observatory (\textit{Swift}), and new missions such as the ECLAIRs telescope onboard the Space Variable Objects Monitor (SVOM) and the Einstein Probe, which detect GRBs of order $\sim$50–80 per year. The second class is represented by all-sky or nearly all-sky MeV monitors, such as the Gamma-ray Burst Monitor (GBM), onboard the {\textit{Fermi}} satellite, which detects GRBs at a significantly higher rate (approximately a factor of three more than narrow-field instruments) but provides poor sky localization, typically spanning tens to hundreds of square degrees. Given that GRBs are predominantly high-redshift phenomena and that VHE emission is strongly attenuated by EBL, only a small fraction of events are expected to be detectable at TeV energies. Combined with the limited duty cycle of IACTs, this implies that maximizing the number of GRBs followed up is crucial for increasing the chances of detecting VHE afterglow emission.\par
\begin{figure}
    \centering
    \includegraphics[width=\columnwidth]{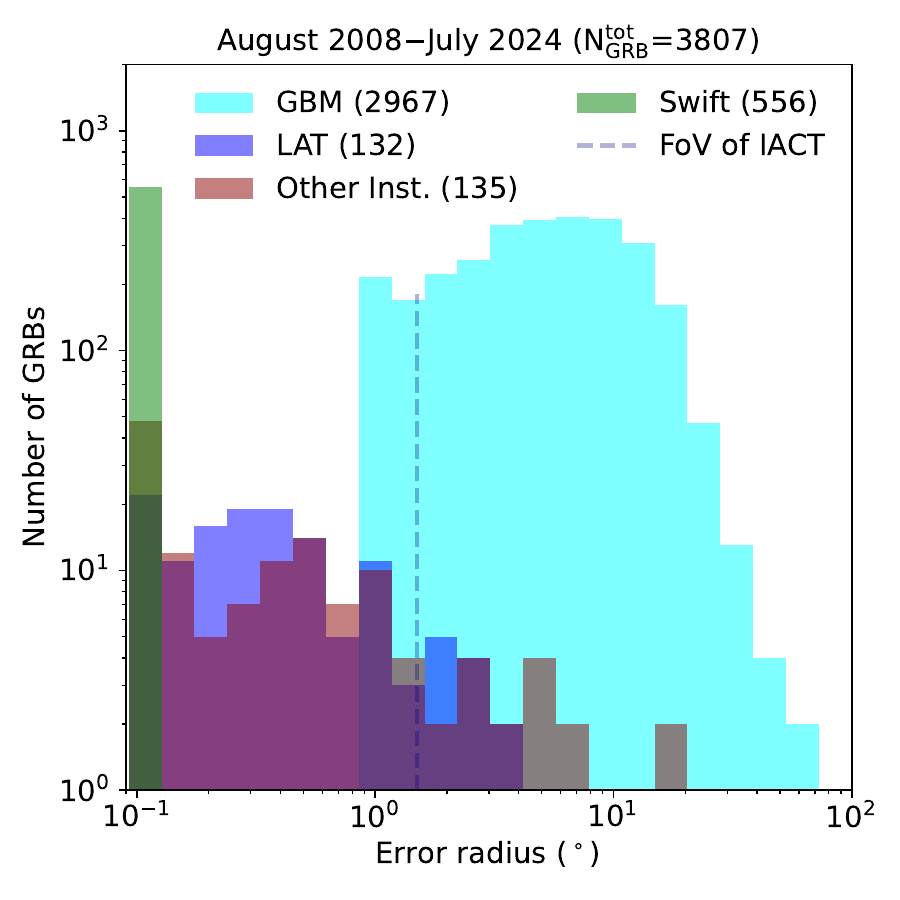}
    \caption{Information retrieved from \url{https://heasarc.gsfc.nasa.gov/w3browse/fermi/fermigtrig.html}. {The instruments which localized GRBs that are detected by GBM are {mentioned in the legend}. In total 2967 GRBs are localized only by GBM}.  Sources with LAT err-radius more than 5$^{\circ}$ excluded (17/3807; counted until 14 July, 2024; 16 years since the launch of \textit{Fermi}). GRBs localized with MAXI, MASTER, AGILE, INTEGRAL, and IPN are included in the category "Other Instruments". The FoV of IACTs is expressed {as radii}. Error {radii} less than 0.1$^\circ$ are considered to be 0.1$^{\circ}$ for plotting.}
    \label{fig:GRB_pop}
\end{figure}

\begin{table}[ht]
  \centering
  \caption{{Typical localization accuracy achieved by the main
           instruments used for GRB detection during August 2008 --
           July 2024 (16 years). Localization errors are quoted as
           typical 90\% containment radii. The symbol "${\dagger}$" denotes the subset of GBM detected GRBs which are localized by XRT or LAT.}}
  \label{tab:localization_accuracy}
  \setlength{\tabcolsep}{10pt}
  \renewcommand{\arraystretch}{1.4}
  \begin{tabular}{lcc}
    \toprule
    {Instrument}
      & {Error radius}
      & {GRBs detected} \\
    \midrule
    \textit{Fermi}/GBM   & $>1^{\circ}$ & $>$2900 \\
    \textit{Fermi}/LAT   & $\sim$$1^{\circ}$           & 149$^{\dagger}$     \\
    \textit{Swift}/BAT   & $1'$--$3'$                  & 1271    \\
    \textit{Swift}/XRT   & $1''$--$5''$                & 556$^{\dagger}$     \\
    \bottomrule
  \end{tabular}
\end{table}
During the period between August 2008 and July 2024 (16 years), \textit{Fermi}/GBM detected more than 3800 GRBs ($\sim$238~GRBs/year), of which more than 75\% ($\sim$2967 GRBs) were localized by GBM alone, with error radii ranging from $1^{\circ}$ to more than $10^{\circ}$. Among those, $\sim$40\% ($\sim$1500~GRBs; $>$80~GRBs/year) are localized within $5^{\circ}$. Over the same period, \textit{Swift}/BAT detected 1271~GRBs ($\sim$79~GRBs/year), approximately half of which ($\sim$35/year) were also observed by GBM. Swift localized a total of 556 GBM-triggered GRB, either through joint detection with BAT (e.g., \ GRB~190114C) or through subsequent \textit{Swift} / XRT follow-up (e.g., \ GRB~230812B). Additionally, Fermi/LAT also detected 149 GBM-triggered GRBs, providing a localization precision of $\sim$$1^{\circ}$, and 132 of these were localized within $5^{\circ}$. A summary of the typical localization accuracy achieved by different instruments is reported in Table~\ref{tab:localization_accuracy}. Mostly, GRBs detected by \textit{Swift}/BAT are followed by ground-based IACTs such as MAGIC and H.E.S.S\@. Given a typical duty cycle of $\sim$10\%, the number of GRBs triggered by Swift/BAT that are potentially observable by IACTs reduces to $\sim$7--8~GRBs/ year, despite GBM triggering a factor of three more GRBs than BAT. Figure \ref{fig:GRB_pop} shows the distribution of sky-localization (error-radius) of GBM-triggered GRBs which are localized by different instruments.

To address these challenges that limit the number of detections, it is critical to develop optimized observational strategies. In this work, we simulate a realistic population of detected GRBs informed by more than a decade of \textit{Fermi}/GBM and \textit{Swift}/XRT observations, including distribution of prompt fluence, isotropic energy, localization accuracy, redshift, and X-ray flux. We investigate how rapid tiling of large sky region can maximize the probability of capturing early TeV emission. Our results provide quantitative examples for the VHE detection rates. We use the observation-scheduling tool, \texttt{tilepy} \citep{Seglar_Arroyo_2024}, {together with \texttt{sensipy} \citep{Green:2026pff}} for a realistic estimate of the GRBs that can be detected following the proposed method.

The paper is organized as follows: in Section 2, we discuss the methodology for constructing a synthetic GRB population based on gamma-ray bursts observed by \textit{Fermi} and \textit{Swift} over more than 15 years. The synthetic population is designed to reproduce realistic observational properties, including coarse localization from GBM (with localization information updated in real time), accurate localization from \textit{Swift}/XRT, and prompt properties such as fluence (S$_{\gamma}$) and the isotropic-equivalent energy (E$^{\gamma}_{\rm iso}$). In Section 3, we discuss the simulated TeV light curve of the GRB in different energy bands. The detectors that are relevant for our studies are discussed in Sect. 4. The main results are presented in Sect. 5 and the observational strategy in Sect. 6. We discuss the detection horizon of the TeV emission in Sect. 7, and {we draw the conclusions} of our study in Sect. 8. 

\section{Methods}

\subsection{Basic assumptions}
To construct a synthetic population of TeV afterglows for a dedicated GRB-triggering instrument, two main ingredients are required. First, a phenomenological model describing the TeV emission must be defined. We adopt the simplest and most commonly used framework for GRB afterglows, namely synchrotron self-Compton (SSC) emission from non-thermal electrons accelerated at the relativistic forward shock formed by the interaction of the jet with the circumburst medium. This model relies on several assumptions concerning the density profile of the circumburst environment, the microphysical parameters of the shock, and the energy distribution of the accelerated electrons.

By construction, this approach neglects several potentially important physical processes, including magnetic-field inhomogeneities in the downstream region \citep{Khangulyan2023,Groselj2024}, self-regulation of the shock due to electron–positron pair production \citep{Derishev2016,Derishev2019}, early pair-loading effects \citep{Thompson2000} in the circumburst medium in the initial phase of the afterglow \citep{Meszaros2001,Kumar2004,Beloborodov2005,RR2007,Ghisellini2010,Nava2013}. {In building} a simple synthetic population of long-GRB TeV afterglows, we ignore these effects and assume a simple adiabatic forward shock propagating into a wind-like medium. Furthermore, we assume that the normalization of the wind density profile is similar for all long-GRB progenitors, that the shock microphysics is universal across the population, that all bursts are observed on-axis, and that the prompt-emission efficiency is the same for all events. Under these simplifying assumptions, the observed VHE afterglow light curve depends only on the isotropic-equivalent energy released in the MeV prompt emission, $E_{\rm iso}$, and on the source redshift, $z$.

The second ingredient is to build the GRB population observed by a given triggering instrument. MeV instruments differ in duty cycle, effective area, and localization cadence and accuracy. Given the large statistical sample accumulated by \textit{Fermi}/GBM over nearly 16 years of operation, we focus on this instrument. Several studies have inferred the intrinsic population of long-duration GRBs in terms of their luminosity function, cosmic formation rate and the jet properties  \citep{Firmani2004,Wanderman2010,Salvaterra2012,Pescalli2015,Petrosian2015,Salafia2015,Palmerio2021,Ghirlanda2022}. VHE afterglows are strongly affected by attenuation due to EBL-absorption, particularly above $\sim200$~GeV, {and above redshift 0.3}. As a result, population models relevant for VHE studies must accurately predict the rate of low-redshift events, which are expected to dominate the observable TeV afterglow population.  We further note that modeling the low-redshift GRB population is particularly challenging. Slightly off-axis jets are likely to contribute at low redshift, but our understanding of prompt emission outside the jet core remains limited. Moreover, the observed sample of GRBs with measured redshifts is biased, due to the reduced efficiency of detecting and localizing faint bursts - peak flux typically below $\sim4~\mathrm{ph~cm^{-2}~s^{-1}}$. To mitigate these limitations, the model dependence of intrinsic low-$z$ GRB populations and the biases affecting observed redshift samples—we adopt the following strategy. We first construct an empirical $(E_{\rm iso}, z)$ distribution based on the observed population of \emph{Swift}- and \emph{Fermi}-detected GRBs, which provides an estimate of the population of observable VHE afterglows in one year of \emph{Fermi} operation.

\subsection{Sample selection}\label{sec:sample_definition}

In order to investigate the proposed strategy of scanning the localization of GBM-detected GRBs, we collected a sample of GRBs detected by \textit{Fermi}/GBM, and a sample of GRBs with measured redshift. The GRBs with known redshift information are available in the Greiner catalog\footnote{\url{https://www.mpe.mpg.de/~jcg/grbgen.html}}. The Greiner catalog is a compilation of GRBs with measured redshift which is regularly updated. It includes well-localized GRBs for which redshifts have been determined from optical afterglows, host galaxies, or other spectroscopic observations. It currently contains about 697 GRBs, which in this study represent the population of GRB named $P_{1}$. In order to obtain precise sky localization information, we also built a catalog of GRBs detected simultaneously by \textit{Swift}/XRT and \textit{Fermi}/GBM. We discuss the details of the observables obtained from XRT and GBM in the following sub-sections. 
\begin{figure}[h]
    \centering
    \includegraphics[width=\columnwidth, height=12cm]{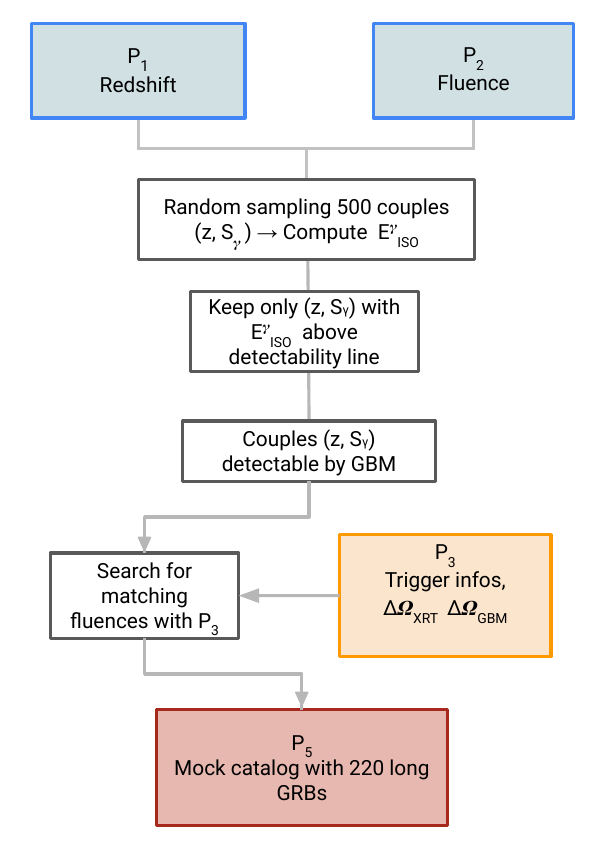}
    \caption{Flowchart for preparation of the GRB catalog. See Sect. \ref{sec:GRBpop} for details.}
    \label{fig:flowchart}
\end{figure}

\subsubsection{\textit{Fermi}/GBM}

GBM triggers on about $\sim$240 GRBs per year, delivering rapid onboard and ground refined localizations distributed through GCN notices and circulars. We collected the fluences (S$_{\gamma}$) of the \textit{Fermi}/GBM detected GRBs up to July 2024 from the GBM catalog\footnote{\url{https://heasarc.gsfc.nasa.gov/w3browse/fermi/fermigtrig.html}}. The resulting distribution, which includes {3807} GRBs, is represented by the skyblue histogram in Fig.\ref{fig:GRB_pop}. 
This population of \textit{Fermi}/GBM GRBs, named in the following $P_{2}$, peaks at $S_{\gamma}\sim 3 \times 10^{-6}~\mathrm{erg\ cm^{-2}}$. 
In addition, the GBM observations provide information regarding the isotropic-equivalent prompt energy E$\rm^{\gamma}_{iso}$, coarse GBM sky-localization, trigger time, the exact arriving time and information from the GCN notices/circulars.

\begin{table*}
    \centering
    \begin{tabular}{clrcc} \hline
        Population & Characteristics       & Number   &  Product & Source  \\ \hline
        \texttt{P$_{1}$} & GRBs with redshift    & 697 &  Redshift distribution & Grainer catalog  \\
        P$_{2}$    & GBM detected GRBs     &  3807    &  S$_{\gamma}$, Trigger info., $\rm\Delta\Omega_{GBM}$         &  GBM catalog       \\
        P$_{3}$    & GRBs (XRT+GBM)        & $370$&S$_{\gamma}$, Trigger info., $\rm\Delta\Omega_{GBM}$, $\rm\Delta\Omega_{XRT}$, F$\rm_{0.3-10 keV}$(t$\rm_{obs}$)& --        \\
        P$_{4}$    & P$_{3}$ with redshift &  110 &   Redshift distribution     &     --    \\ \hline
        \multirow{2}{*}{P$_{5}$}    & \multirow{2}{*}{Mock catalog}          &    \multirow{2}{*}{220}   &  S$_{\gamma}$, E$\rm^{\gamma}_{iso}$, Trigger info., &   --      \\
                   &                       &&  $\rm\Delta\Omega_{GBM}$, $\rm\Delta\Omega_{XRT}$, F$\rm_{0.3-10 keV}$(t$\rm_{obs}$)&         \\ \hline        
    \end{tabular}
    \caption{populations of GRBs that lead to the mock catalog of GRBs. See Sect. \ref{sec:GRBpop} for details.}
    \label{tab:Catalog}
\end{table*}

\subsubsection{\textit{Swift}/XRT}\label{sec:xrt}

For the purpose of this study, we selected a sample of GRBs with simultaneous detection with XRT and GBM, which defines the population $P_{3}$. In particular, from this joint population 
it is possible to access the afterglow X-ray light curves in 0.3-10 keV and high-precision (arc-second) localization. A subset of 110 GRBs in this joint sample has measured redshifts, taken from the Greiner catalog. 
To investigate common behaviors in the X-ray afterglows of GRBs, we analyzed the temporal evolution of the XRT light curves (LCs) for the GRBs in our joint XRT--GBM sample. $P_{4}$ is a subset of GRBs in $P_{3}$ with a measured redshift, used for characterizing the temporal decline of X-ray light curves (see Fig.~\ref{fig:XRT_clustering}).
Among the 370 GRBs jointly detected by XRT and GBM, 27 are short GRBs (7--8\% of the sample), and 100 GRBs exhibit flares ($\sim 27\%$ of the sample), consistent with population studies (Tiwari et al in prep.) indicating that roughly one-third of GRBs display X-ray flares. Following \citet{2012MNRAS.425..506D}, we normalized the XRT flux by the GBM fluence. The resulting clustering in normalized flux highlights the general temporal trends in X-ray afterglows and provides a framework to compare our joint XRT--GBM sample with previous studies. 

\subsection{Synthetic population of GRBs}\label{sec:GRBpop}

To generate a synthetic GRB population, we begin with two observed datasets: the complete \textit{Fermi}/GBM GRB population $P_{2}$ and the full redshift catalog compiled by Greiner $P_{1}$. 
These serve as empirical priors for sampling the prompt emission properties of the mock population. Our goal is to produce a sample of long GRBs that is representative of one year of \textit{Fermi}/GBM detections, while retaining realistic properties such as fluence, redshift, $E^{\gamma}_{\mathrm{iso}}$ and sky localization.  

The procedure is illustrated in the workflow shown in Fig.~\ref{fig:flowchart}. We randomly extracted 500 fluences from $P_{2}$ and 500 redshifts from $P_{1}$. For each pair $(S_{\gamma}, z)$, we computed the isotropic equivalent gamma-ray energy as 
\begin{equation}
\rm 
    E^{\gamma}_{\mathrm{iso}} = \frac{4 \pi d_{L}^{2}}{1+z}S_{\gamma}
    \label{E_iso}
\end{equation}
and applied a detectability cut based on the $E^{\gamma}_{\mathrm{iso}}$--$z$ relation in Fig.~18 of \citet{2019ApJ...878...52A}. Only pairs above the detectability line were retained.

Next, to create a population corresponding to one year of GBM observations with realistic sky localization, we use population $P_{3}$. For each pair accepted in $(S_{\gamma}, z)$, we search in $P_{3}$ for a real GRB with similar fluence. When a match is found, the simulated GRB inherits the real GRB's GBM trigger time, GCN Notices, GBM sky localization information, and XRT accurate localization. This guaranties realistic sky positions fully consistent with GBM observational constraints. Finally, we randomly selected 220 GRBs from these matches to form the final mock population of a year of long GRBs observed by GBM: $P_{5}$.
 
In summary, each entry in $P_{5}$ includes fluence, $E_{\mathrm{iso}}$, redshift, GBM localization, and XRT localization. 
The matching real GRB also provides its corresponding XRT afterglow light curve, allowing for a direct comparison between the simulated population and the real XRT--GBM sample. 
The redshift and fluence distributions of $P_{5}$ are shown in Fig.~\ref{fig:GRB_catagorization} alongside those of $P_{1}$ and $P_{2}$ respectively, to illustrate the consistency between the simulated and observed population.
In Fig.~\ref{fig:detection_limit} we further demonstrate the consistency of $P_{5}$ by comparing the $E^{\gamma}_{\mathrm{iso}}$ vs $z$ distribution with the GBM detectability limit. A detailed discussion of the $E^{\gamma}_{\mathrm{iso}}$-$z$ relation is provided in \citealt{2023ApJ...952L..42L}.

Although the sample of GRBs with measured redshift is known to be observationally biased toward events with brighter afterglows and successful optical follow-up, this bias does not affect the aims of this work. Our goal is not to reproduce the intrinsic cosmic GRB population, but rather to simulate a realistic scenario of one year of GRB observations of \textit{Fermi}/GBM. For this purpose, an observationally biased redshift distribution reflects the subset of GRBs for which multiwavelentgh follow-up is feasible in practice. Moreover, after pairing each fluence with a redshift from this distribution, we require the event to satisfy the GBM detectability condition derived from the $E^{\gamma}_{\mathrm{iso}}$-$z$ relation. This ensures that only physically plausible GBM detectable GRBs are retained, preventing the bias from introducing unrealistic faint high redshift events.

\section{Modeling of GRB afterglow}\label{sec:LC_VHE_model}

Following the framework outlined by \citealt{Tiwari:2025tgr} and \citealt{Banerjee:2024hxp}, we specifically utilized the leptonic component of the Lepto-Hadronic Modeling Code~\citep[\texttt{LeHaMoC};][]{Stathopoulos:2023qoy}, \texttt{LeMoC}\footnote{\texttt{LeMoC} (\url{https://github.com/mariapetro/LeHaMoC}) enables the calculation of energy losses in leptons due to processes like synchrotron radiation, inverse Compton (IC) scattering, synchrotron self-absorption, adiabatic losses, and photon-photon absorption.}. The spectral shape in afterglow is governed by synchrotron and inverse Compton radiation from power-law distributed electrons accelerated at the forward shock front. Particle injection takes place within a region of radius $R_e$, in a magnetic field of strength B. 

We consider that the proportion of total energy in the shocked region used to accelerate electrons into a power-law distribution ($\epsilon_{\rm e}$) and to enhance magnetic fields ($\epsilon_{\rm B}$) remains constant over time. The electrons injected and accelerated by the shock follows a power-law distribution (denoted by $dN/d\gamma \propto \gamma_e^{-p}$), wherein $p$ and $\gamma$ denote the spectral index of the electrons and the electron Lorentz factor, respectively. We further assume that the index $p$ does not evolve over time. 

Within the synchrotron self-Compton (SSC) framework, previously reported studies on long gamma-ray bursts (long-GRBs) indicate that the parameter set \(p\sim2.3\), \(\epsilon_{\rm e} = 0.1\), \(\epsilon_{\rm B} \sim 10^{-4}\), and a wind-like circumburst medium successfully reproduces the observed simultaneous X-ray and GeV emissions \citep{Tiwari:2025tgr}. It is important to note that both the prompt emission efficiency is taken to be $10\%$. This particular set of parameters is also adopted for predicting the VHE emission in afterglows.

\subsection{EBL attenuation}

The unabsorbed intrinsic spectra of the simulated GRBs are estimated between 20\,GeV-5\,TeV as previously described. We apply further attenuation due to EBL, which absorbs high-energy $\gamma$-ray photons through pair production with low-energy EBL photons along the line of sight. The observed spectra are therefore modified with an exponential suppression term, {$e^{-\tau_{\gamma\gamma}(E,z)}$} , such that 
\begin{equation}
    \rm F _{{E_1},{E_2}} ^{{obs}} = \int_{E_1}^{E_2} E \frac{dN}{dE}\, \exp [-\tau_{\gamma\gamma}(E,z)] ~dE,
\end{equation}

where  {$\tau_{\gamma,\gamma}(E,z)$} is the optical depth predicted by the EBL model proposed by \citealt{2011MNRAS.410.2556D} for a specific redshift \textit{z} and at energy $E$, and $\rm \frac{dN}{dE}$ is the intrinsic photon spectrum derived from the model presented in Sect.~\ref{sec:LC_VHE_model}. The energy bounds E$_{1}$ and E$_{2}$ define the range over which the observed integral flux ($\rm F _{{E_1},{E_2}} ^{{obs}}$) is computed. 

To implement the EBL attenuation in practice, we used the \texttt{TemplateSpectralModel} function provided by \texttt{Gammapy} \citep{gammapy:2023}. For each time step, the intrinsic photon spectrum of the GRB was provided as input to a template spectral model incorporating the EBL attenuation from \citealt{2011MNRAS.410.2556D}. The model automatically evaluates the energy-dependent optical depth and applies the corresponding absorption factor across the entire energy grid. This procedure ensures a consistent and accurate computation of the EBL-suppressed spectra and light curves for all GRBs in our simulated sample.

\section{Observatories}\label{sec:Detectors}
\begin{table*}[ht!]
    \begin{tabular}{lcccccccc} 
    IACT & Band [TeV]          & 20 GeV  & 200 GeV & 1 TeV    & Field of View           & Fastest Slew    & Location          & Duty Cycle \\ \hline 
    CTAO-N & 0.02-300           &\greenbox &\greenbox &\greenbox& $\sim4.5^\circ-8^\circ$             &$\lesssim 30$\,s (LST)  & La Palma (Spain)  & $\sim10\%$ \\ \hline 
   ASTRI & $\sim$0.5-10       & \redbox& \redbox &\greenbox& $10^\circ$ & $\sim60$\,s &  Tenerife (Spain)   & $\sim10\%$ \\ \hline 
    LACT &   $>0.2$    & \redbox&\greenbox&\greenbox&      $\sim10^\circ$          & --      &  Daocheng  (China) & $\sim10\%$    \\ \hline 
    \end{tabular}
    \caption{Summary of the detectors used in this work. The color-bars in coloumn 3, 4, and 5 indicate the accessible energy bands of each detector, where the green- and red-bar indicate the accessible and not-accessible energy bands, respectively. {References for the different observatories listed here are reported in Sect.~\ref{sec:Detectors}.}}
    \label{tab:Detectors}
\end{table*}
For this study, we focus on ground based atmospheric Cherenkov telescopes, such as ASTRI {\citep{ 2024Univ...10...94V,Vercellone:2022ull}}, LACT \citep{Zhang:2025Ho} and CTAO \citep{CTAConsortium:2017dvg} (see Tab.~\ref{tab:Detectors}). 
Throughout the paper, we discuss results based on CTAO-N alpha-configuration that comprises the Large-Sized Telescope (LSTs), the Medium-Sized Telescope (MSTs) and the Small-Sized Telescope (SSTs).
Figure~\ref{fig:Detector_sensitivity} shows the differential flux sensitivity of the telescopes for an integration time of 50 hours. {We note that the energy threshold of LACT is quoted as 250 GeV in \citet{Zhang:2025Ho} for a zenith angle of 20$^{\circ}$. However, for this study, we assumed a simple extrapolation to 200 GeV for comparison with other facilities.}
The short slew times and broad energy coverage make these observatories well suited for rapid follow-up of GRB alerts from \textit{Fermi}/GBM, enabling the detection of the VHE emission. For example, CTAO {will} typically repoint in less than 30 s\footnote{{The 30 s represent the fastest slewing time, which is specifically valid for the LSTs. We note that the slewing time of the MSTs and SSTs can be up to about one minute.}} and provides sensitivity over a wide energy range, from approximately 20~GeV to 300\,TeV. 
Since the field of view of the pointing telescopes (LACT, ASTRI, and CTAO-N) remains limited (diameter of {$4^\circ-10^\circ$}), the well-localized GRB-triggers are typically prioritized. 
The sensitivity of these pointed telescopes is significantly higher in the 0.2-1\,TeV energy range than that of LHAASO, a water Cherenkov particle detector. Nevertheless, LHAASO benefits from an exceptionally high duty cycle ($>$95\%) and a substantially larger field of view (FoV). Thus while LHAASO can monitor a large number of GRBs, achieving a detection significance above 5$\sigma$ remains a challenge, mainly due to limited sensitivity at energies below 1 TeV.
Therefore, in this work, we discuss the capability of pointing-telescopes to scan the patch of the localization of GBM-detected GRBs.

\begin{figure*}[ht]
    \centering
    \includegraphics[width=\columnwidth, height=7cm]{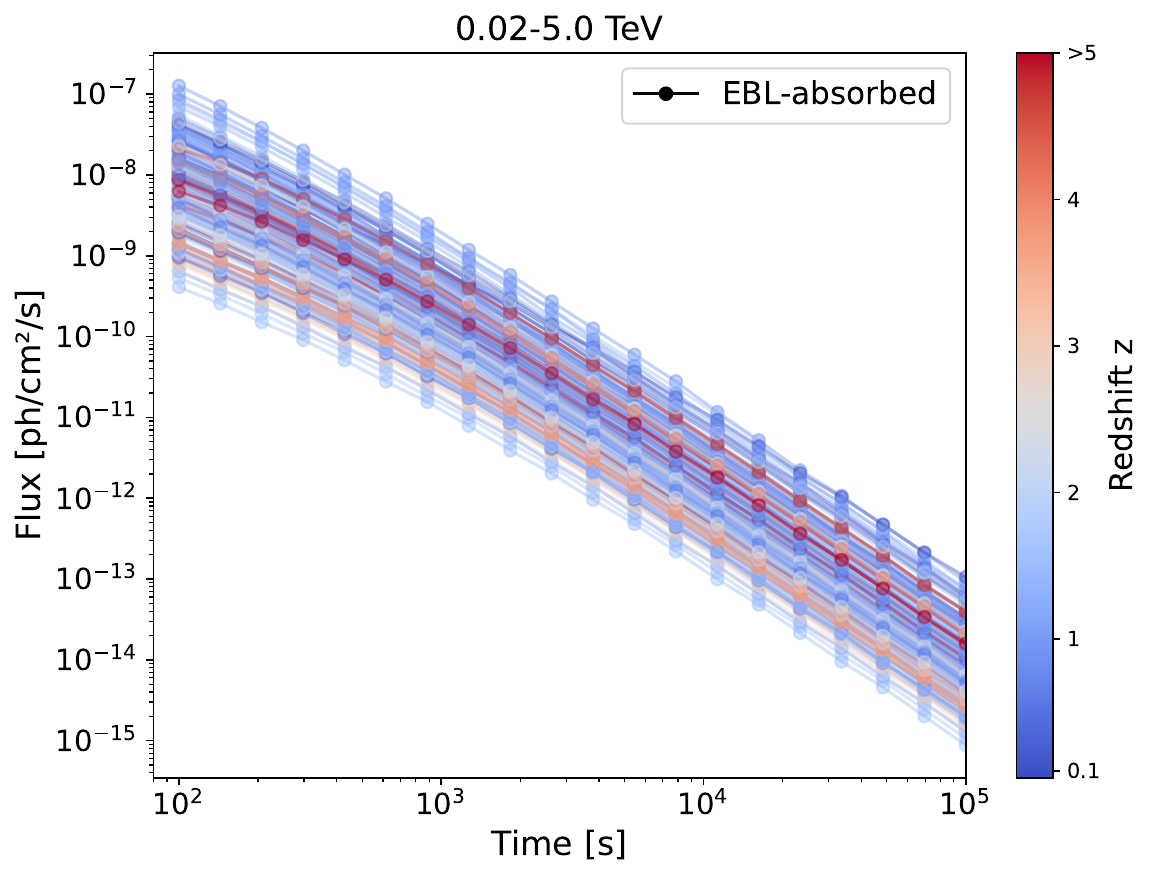}
    \includegraphics[width=\columnwidth, height=7cm]{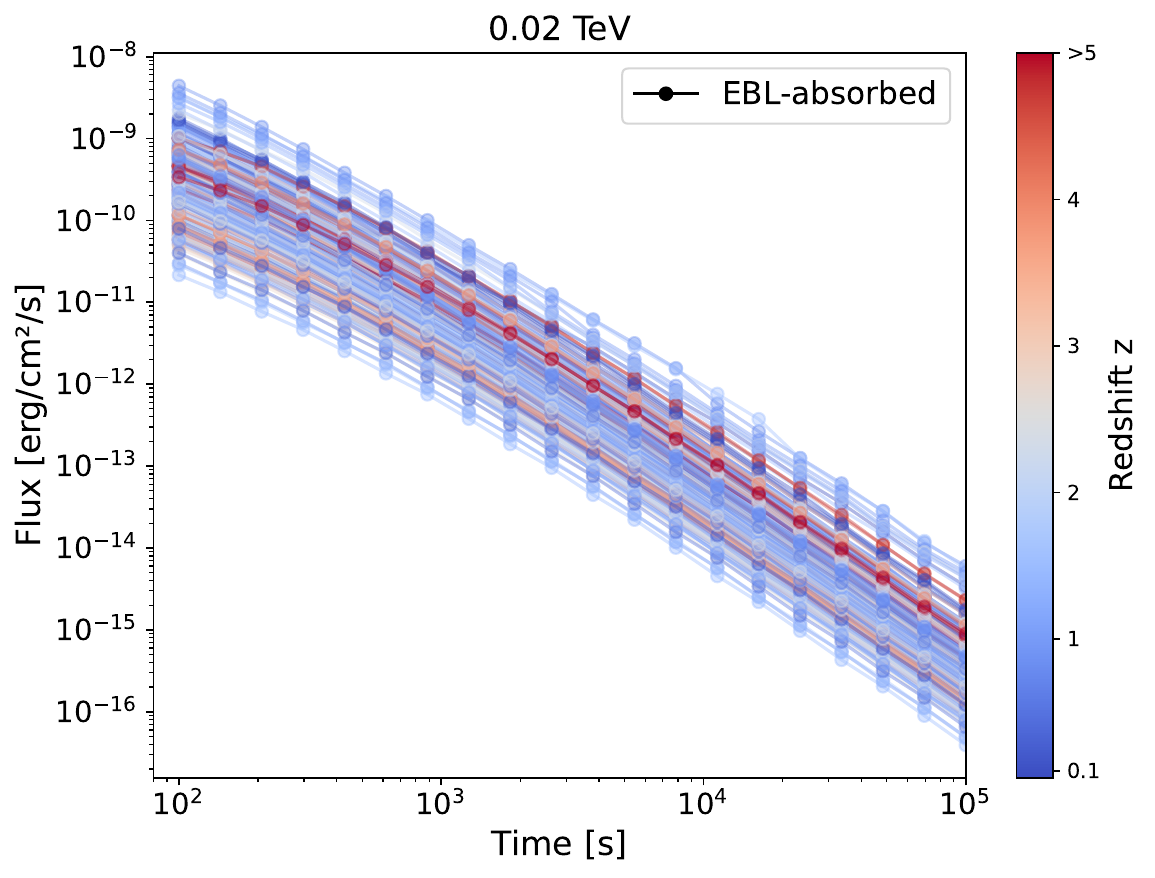}
    \includegraphics[width=\columnwidth, height=7cm]{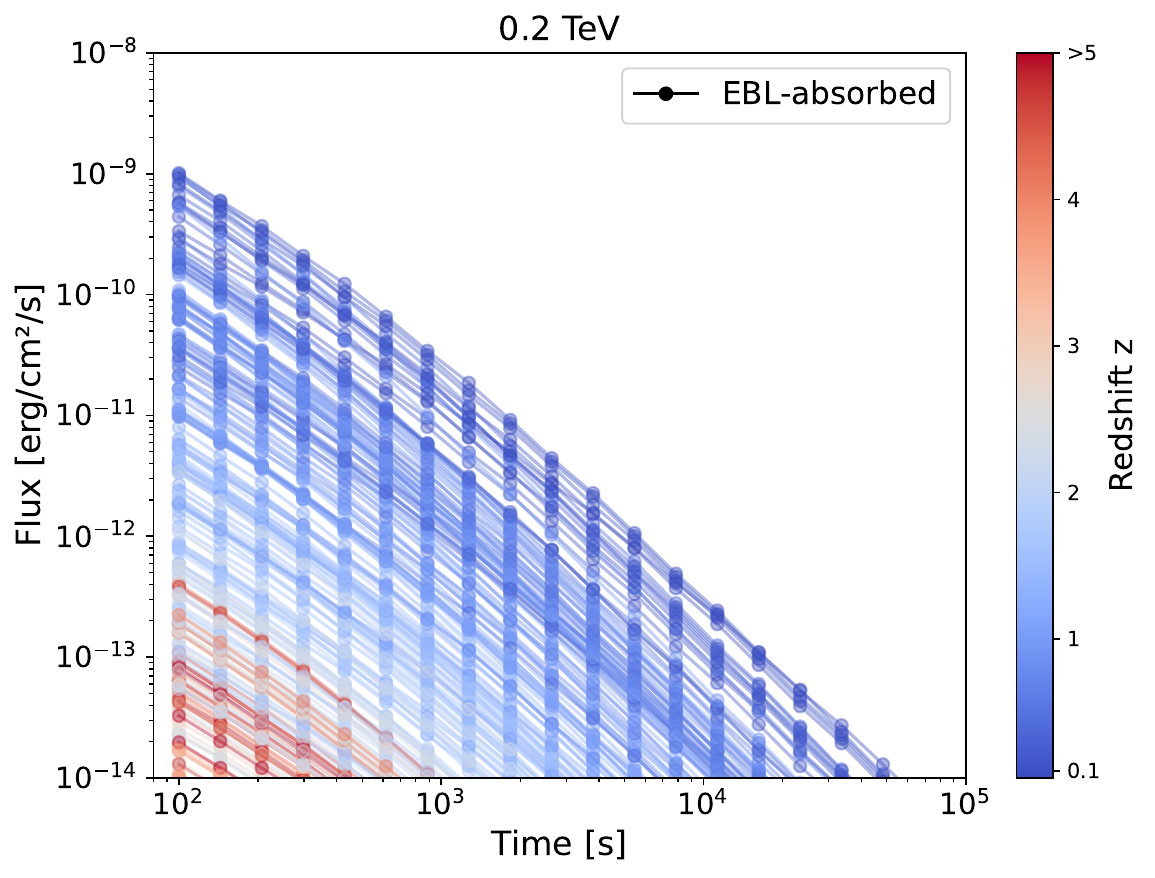}
    \includegraphics[width=\columnwidth, height=7cm]{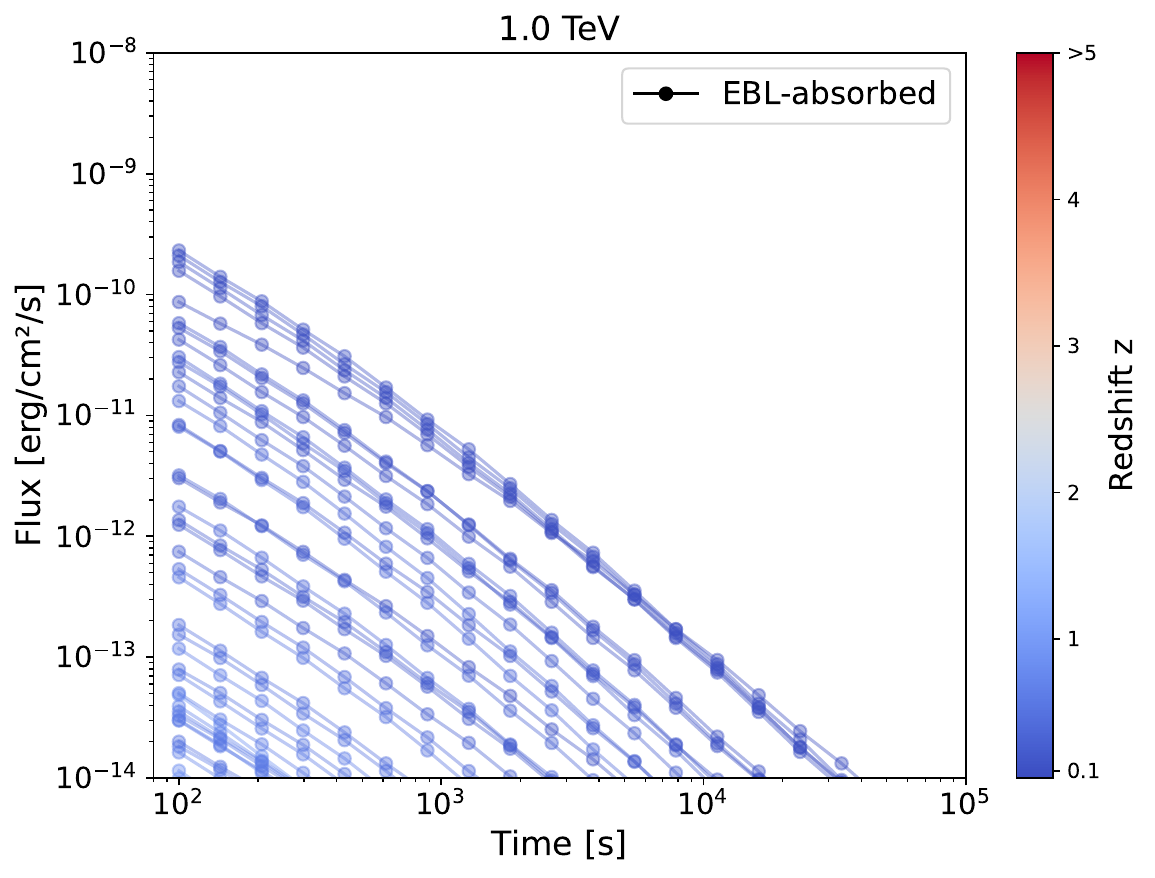}
    \caption{The simulated light curve of the synthetic population of GRBs in 0.02-5\,TeV (energy integrated light curve; top left), 0.02 TeV (top right), 0.2 TeV (bottom left) and 1 TeV (bottom right). The time resolved SEDs are computed at each points on the light curve which are further used for obtaining the EBL absorbed (observed spectra). The color bar on the right indicate the redshift of the GRBs.}
    \label{fig:VHE_LCs}
\end{figure*}

\section{Results}

\subsection{Temporal behavior of simulated TeV emission and observed X-ray data}

Based on the mock-catalog (P$_{5}$, see Sect.~\ref{sec:GRBpop}) and using intrinsic properties of the long GRBs (see Sect. \ref{sec:LC_VHE_model}), we derived the light curves and time-resolved VHE spectra. We consider the energy-integrated light curves in the range 0.02-5 TeV, shown in the top-left panel of Fig.~\ref{fig:VHE_LCs}, as well as light curves evaluated at specific energies: 20 GeV, 200 GeV, and 1 TeV, shown in the other panels of Fig. \ref{fig:VHE_LCs}. In Fig.~\ref{fig:XRT_VHEflux} we show the temporal evolution of the X-ray afterglow light curves (observed with \textit{Swift}/XRT) and the simulated intrinsic VHE light curves, both normalized with the prompt fluence (S$_{\gamma}$) associated to the simulated GRBs. 
We note that, similar to the X-ray emission, the TeV light curves tend to cluster, indicating a relatively uniform scaling with the prompt energetics. Moreover, the TeV and X-ray light curves show a strong correlation in both normalization and temporal behavior during the early afterglow phase: both bands initially decay with a comparable power-law slope, consistent with the expectation that the same population of shock accelerated electrons drives both synchrotron (X-ray) and IC (GeV-TeV) emission. This is expected due to the choice of intrinsic parameters, $\epsilon_{\rm B}\sim$10$^{-4}$, p$=2.3$, the efficiency of prompt $\eta\sim0.1$ and $\epsilon_{\rm e}\sim0.1$ as described in Sect.~\ref{sec:LC_VHE_model} and \citealt{Tiwari:2025tgr}). In the Table~\ref{tab:ssc_params_comparison} we compare the microphysical parameters inferred from the modelling of VHE-detected afterglows by different authors.   

Interestingly, at later times (after 3\,ks), the TeV flux declines rather rapidly than the X-ray emission, which has a shallower decay. This divergence can reflect different factors, including decreasing efficiency of IC emission as the blast wave slows down, and it was also observed in some of the TeV detected GRBs, such as GRB 190829A \citep{2021Sci...372.1081H}, for which an evident difference in the flux levels in X-ray and TeV has been observed\footnote{Although the temporal decay indices in X-rays and VHE $\gamma-$rays are similar, the ratio between the fluxes at $\sim$20\,ks is a factor of three less in VHE $\gamma$-rays (0.2-4\,TeV) as compared to the X-rays. See \citealt{2021Sci...372.1081H} for details.}. We further note that the afterglow model used in this work is based on the synchrotron self-Compton scenario, which does not include additional physical processes, such as hadronic contributions, evolving microphysical parameters\footnote{For example, the pair balance model, See \citealt{DP2021} for details.}, or delayed energy injection) which might impact the TeV light curves at later times. Therefore, the faster decline of the TeV flux at later epochs may also reflect the limitations of our simplified approach.

\begin{figure}[h]
    \centering
    \includegraphics[width=\columnwidth, height=7cm]{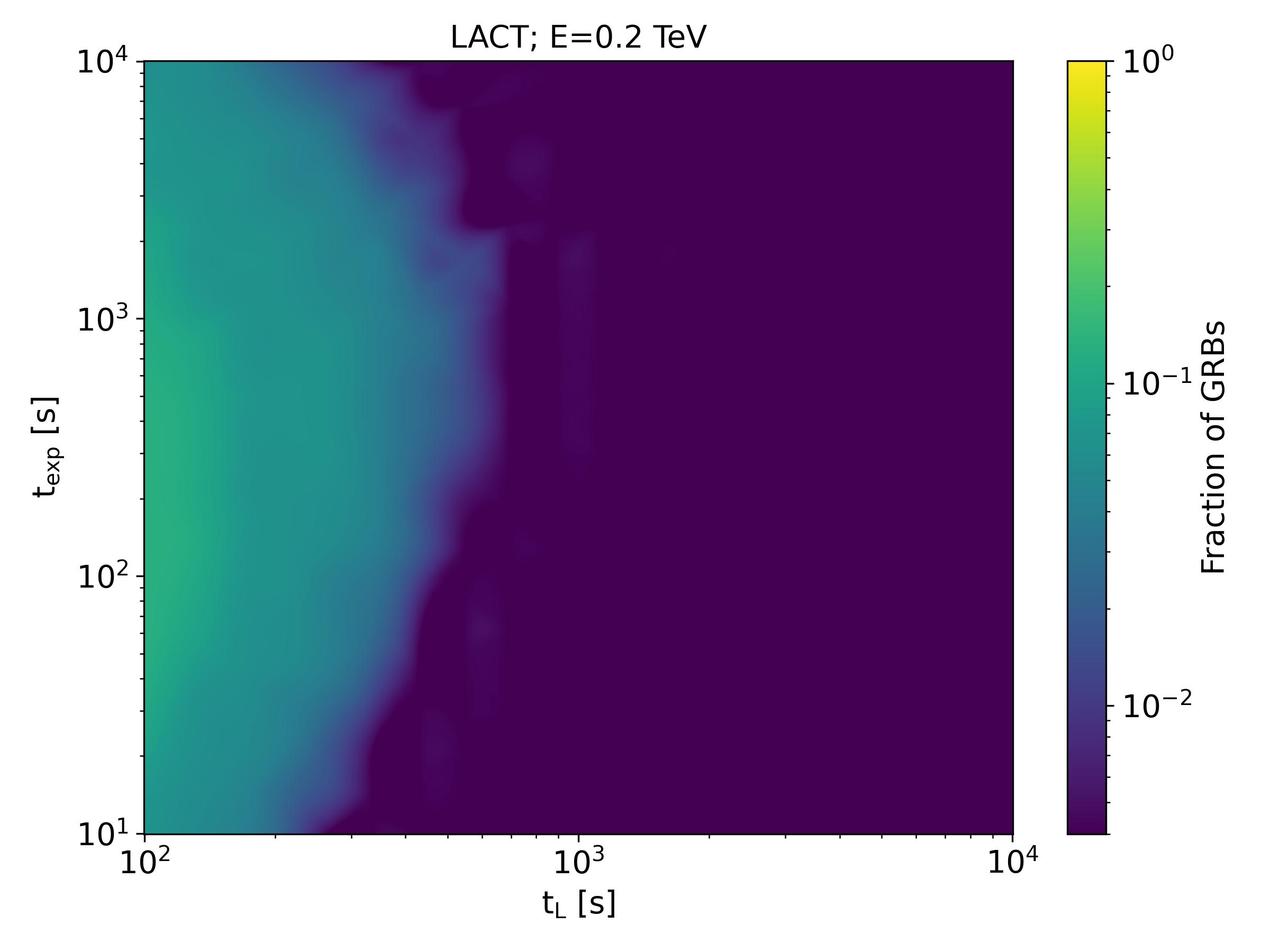}    \includegraphics[width=\columnwidth, height=7cm]{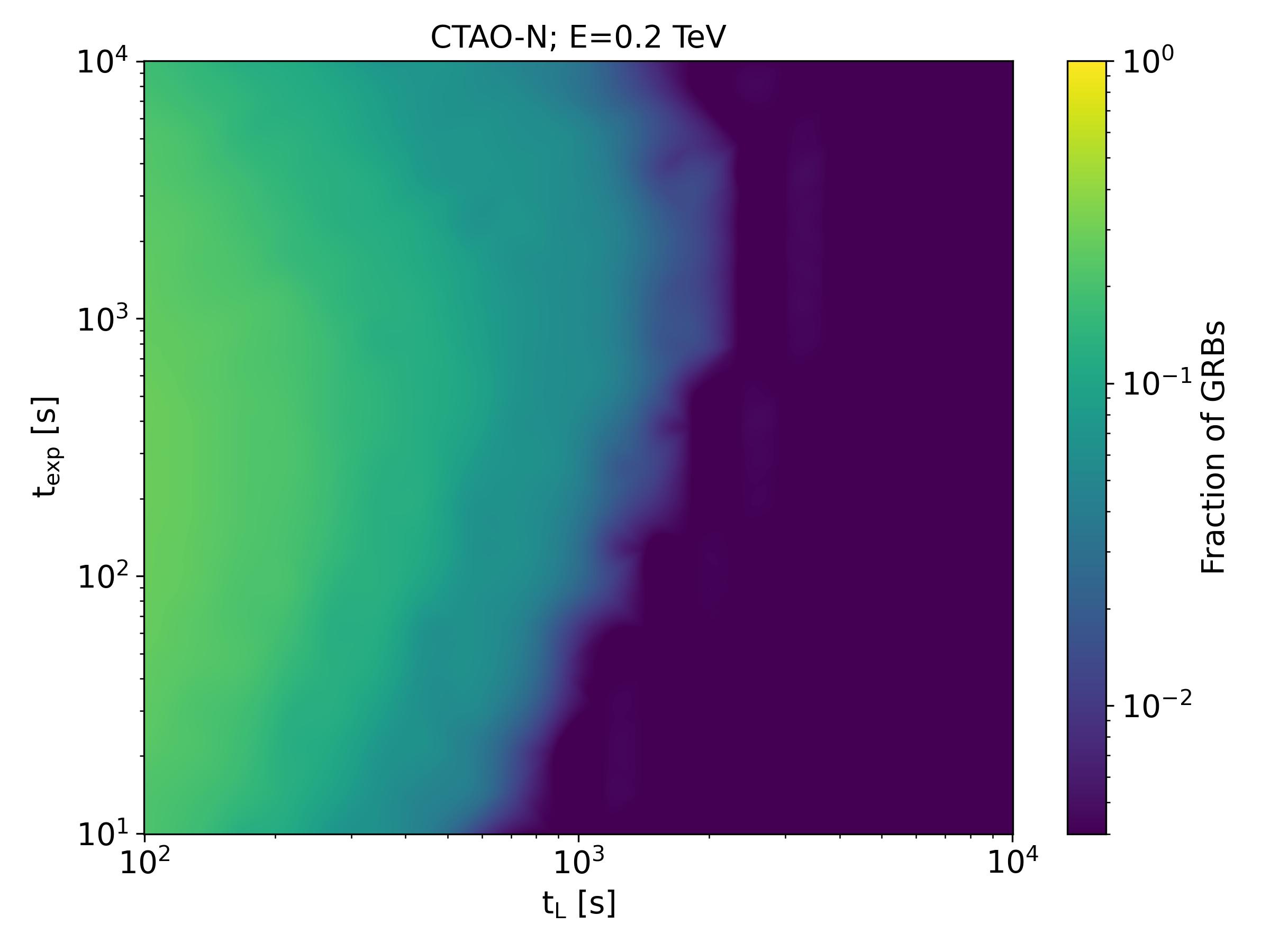}
    \caption{Heatmap showing the detectability of the GRBs in the mock catalog at 200 GeV for LACT and CTAO-N. The latency (t$\rm_{L}$) and the exposure time (t$\rm_{exp}$) of the observations are indicated in the horizontal and the vertical axes. The color-bar indicates the fraction of GRBs detectable for a combination of the latency time and the exposure time.}
    \label{fig:LACT_200GeV_dl_exp}
\end{figure}

\begin{figure}[h]
    \centering
    \includegraphics[width=\columnwidth, height=7cm]{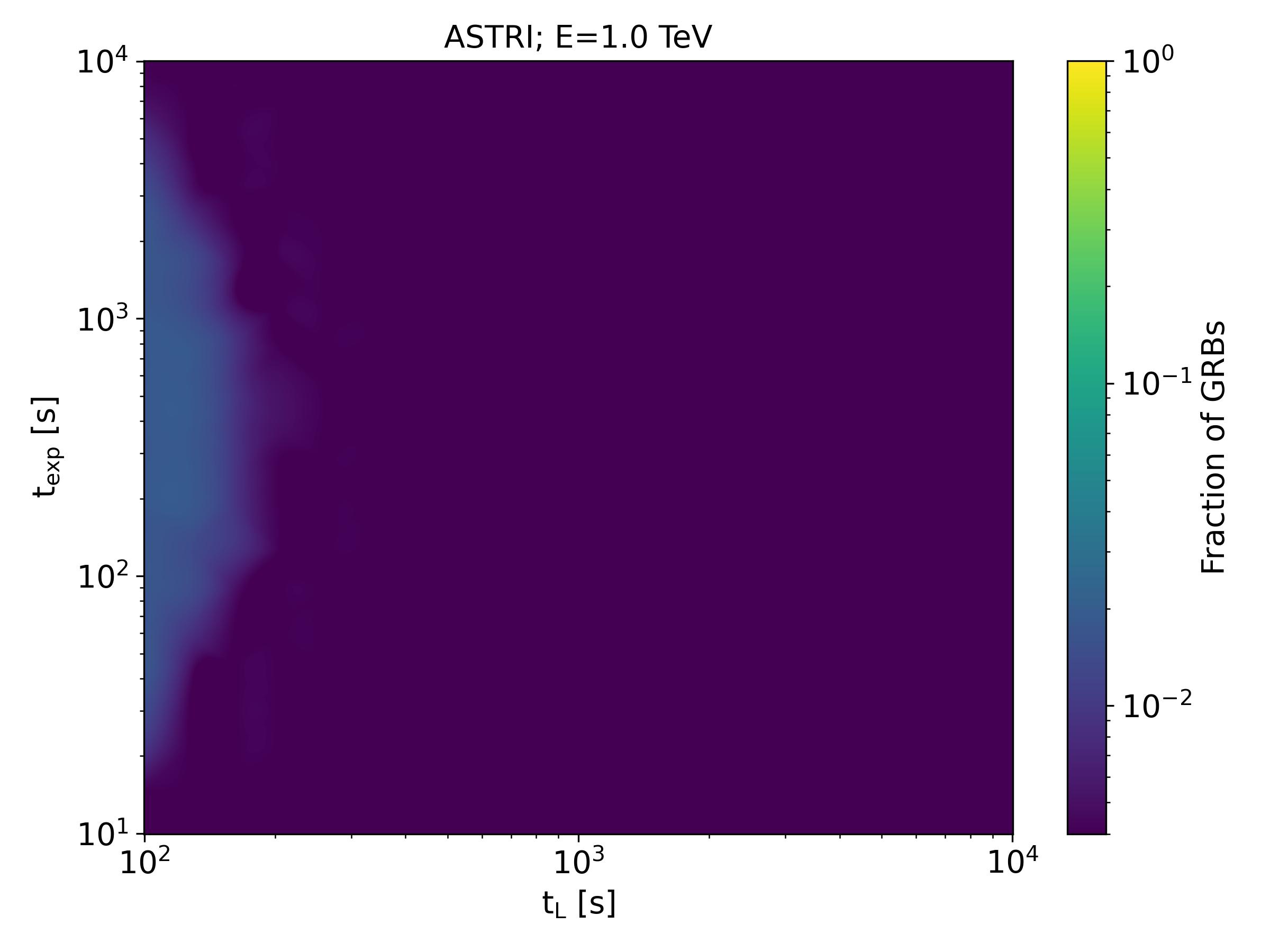}
    \includegraphics[width=\columnwidth, height=7cm]{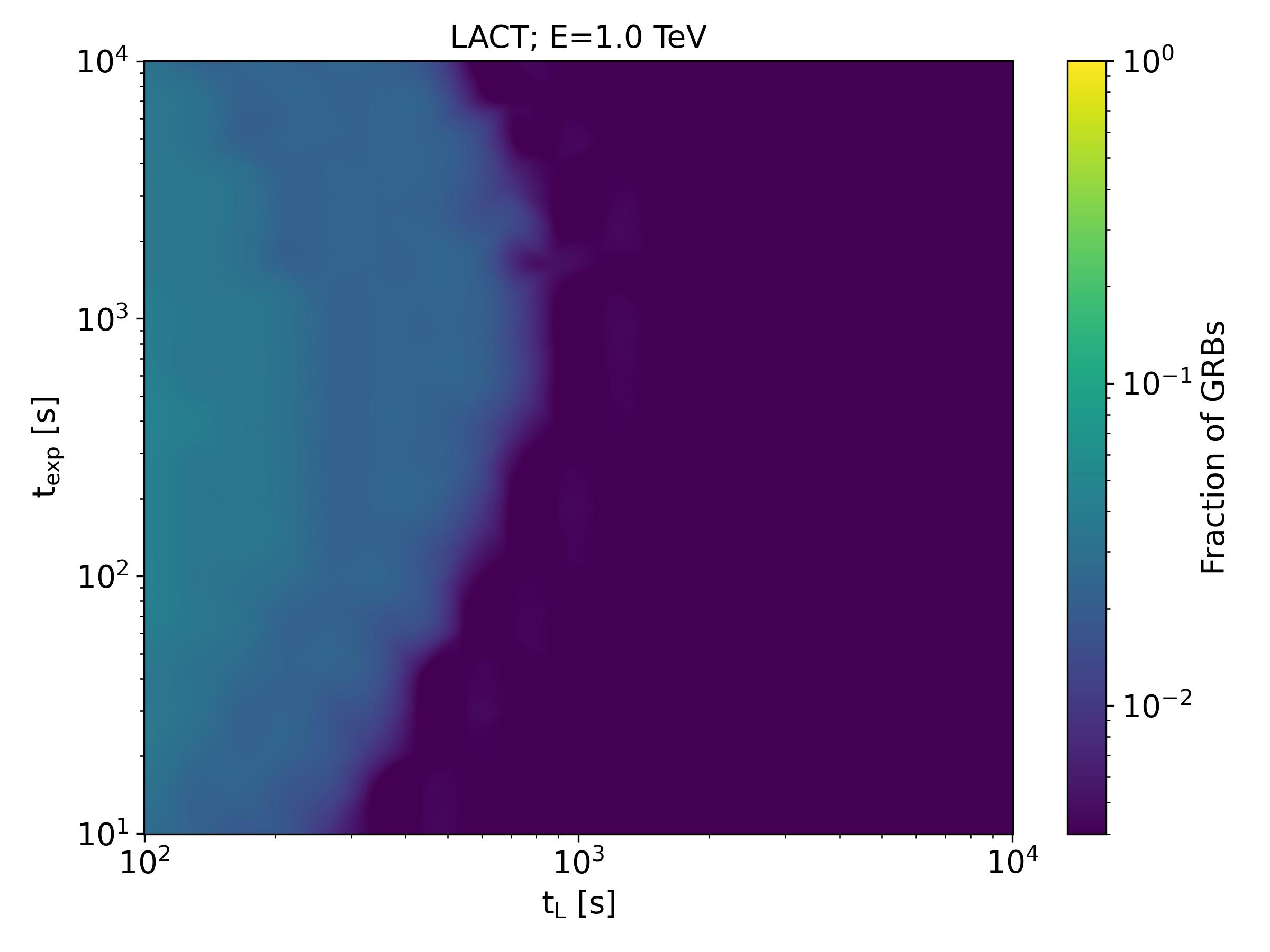}
    \includegraphics[width=\columnwidth, height=7cm]{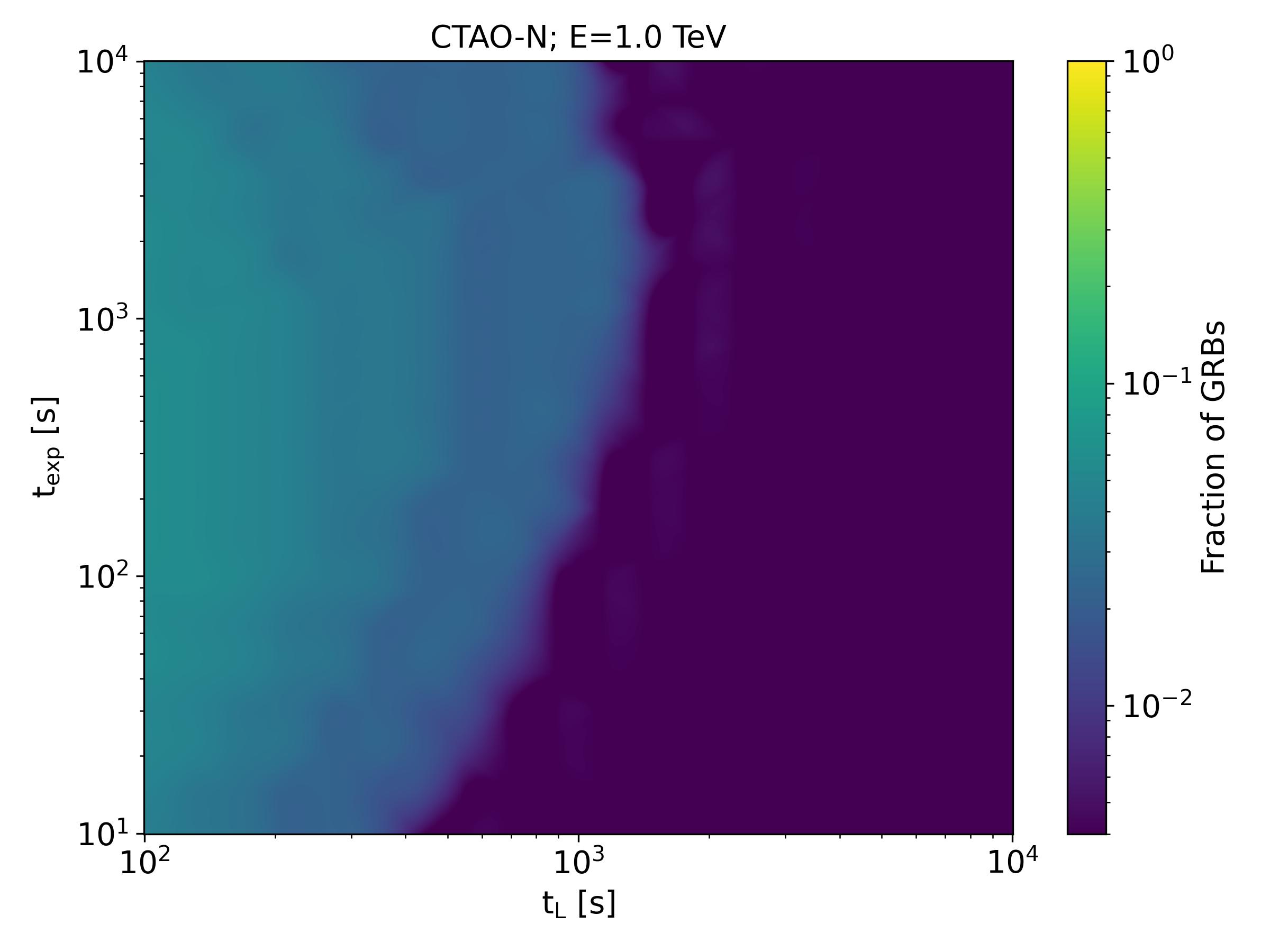}

    \caption{Same as Fig. \ref{fig:LACT_200GeV_dl_exp} for ASTRI, LACT and CTAO-N.}
    \label{fig:1TeV_dl_exp}
\end{figure}

\subsection{Fraction of detectable GRBs}

We convolved information on the sensitivity of TeV detectors, such as LACT, ASTRI and CTAO-N; see Fig. {\ref{fig:Detector_sensitivity}}) with the 
simulated light curves of GRBs (see Sect. \ref{sec:LC_VHE_model}) from the
population of GRB (see Sect. \ref{sec:GRBpop}). The minimal latency of observation has been set to 100 \,s (T$^{\rm obs}_0$). Here, we present two different methods to estimate the detectability of GRBs:\par
a) Absence of a publicly available instrument response function (IRF). This is true for the instruments 
ASTRI and LACT. In this case, we used the differential sensitivity as presented in Fig.~\ref{fig:Detector_sensitivity} and scaled the limiting flux assuming that the sensitivity scales are t$\rm _{exp}^{-0.5}$ (see Eq.~\ref{eq:Flim}), the exposure time of the observation. \par

b) For CTAO-N, since IRFs are publicly available \citep{cherenkov_telescope_array_observatory_2021_5499840}\footnote{\url{https://zenodo.org/records/5163273}}, we used the public simulation tools \texttt{sensipy} \citep{Green:2026pff} and \texttt{tilepy} \citep{Seglar_Arroyo_2024} to estimate the detectability and design a realistic observational strategy for CTAO-N energy range. The details of the procedure are presented in Sect. \ref{sec:observation-strategy}.\par
  
In the following, we describe a generic approach to obtain the number of detectable GRBs based on differential sensitivity.

\subsubsection{Using differential sensitivity}\label{sec:diffSen}
We have considered the flux limits of the sensitivity curves as shown in Fig. \ref{fig:Detector_sensitivity}. The differential flux sensitivity (F$^{\rm \Delta T}_{\rm lim}$) for a shorter exposure of $\rm \Delta T$ is calculated using the following equation.
\begin{equation}\label{eq:Flim}
    \rm F^{\rm t_{exp}}_{\rm lim}(\rm E)=  F^{\rm \Delta T}_{\rm lim}(\rm E) \times  \left( \frac{t_{exp}}{\Delta T}\right)^{-0.5},
\end{equation}

where $\rm F^{\rm \Delta T}_{\rm lim}(\rm E)$ is the flux sensitivity bound from Fig. \ref{fig:Detector_sensitivity} at the energy E for an observation period of $\rm \Delta T=50$\,hrs. We assumed that the sensitivity varies as t$^{-1/2}$, where t is effective {ON}\footnote{The ON time is defined as the total effective exposure time after applying the data-quality cuts.} time.  $\rm F^{\rm \Delta t'}_{\rm lim}(\rm E)$ denotes the flux sensitivity for the on observation time of $\rm \Delta t'$. 

We compare the flux for each GRB (F$\rm ^{t_{obs}}_{E, model}$ as derived from the modeling described in Sect. \ref{sec:LC_VHE_model}) in our sample (see Sect. \ref{sec:GRBpop}) at energy E (20 GeV, 200 GeV and 1 TeV) at a time $\rm t_{obs} = \sqrt{\rm t_{L}\times(t_{L}+t_{exp})}$ with the sensitivity limit, F$^{\rm t_{exp}}_{\rm lim}(\rm E)$, with exposure t$_{\rm exp}$ at a specific energy E. We consider a detection if  F$\rm ^{t_{obs}}_{E, model}>$ F$^{\rm t_{exp}}_{\rm lim}(\rm E)$ is satisfied. We simulate this detection scenarios with a distribution of latency time (t$\rm _{L}$; $\rm 10^{2}s-10^{4}s$) and exposure time (on source time: t$\rm _{exp}$; $\rm 10^{1}s-10^{4}s$) at two different energies: 200 GeV and 1 TeV. 

We consider two energy bins: 200 GeV and 1 TeV. This selection is based on the energy threshold of the IACTs depending on the zenith angles. The energies 200 GeV and 1 TeV correspond to the low (5$^{\circ}$-35$^{\circ}$) and/or medium (35$^{\circ}$-50$^{\circ}$) and high-zenith ($>50^{\circ}$) cases, respectively. 
Through this method we discuss the detectability of sources assuming that they are followed-up with a latency of t$_{L}$ without a duty factor of the IACTs. Fig.~\ref{fig:LACT_200GeV_dl_exp} and Fig.~\ref{fig:1TeV_dl_exp} present the fraction of detectable GRBs as a function of the latency time and the exposure time for the two energies 200 GeV and 1 TeV, and for the different observatories considered in this study. 

\section{Observation strategy}
\label{sec:observation-strategy}

\subsection{Localization Information}

For each GRB 
of our mock sample P$_{5}$ (see Tab.\ref{tab:Catalog}), we provide the following information:
fluence (S$_\gamma$), isotropic equivalent energy in prompt (E$\rm^{\gamma}_{ISO}$), sky-localization from \textit{Swift}/XRT (arc-minute accuracy; $\rm\Delta\Omega_{XRT}$) along with trigger information such as updates on the sky-localization information from GCNs from GBM triggers ($\rm\Delta\Omega_{GBM}$) and the arrival time of the notices. Right after a discovery of a GRB, three different notices are sent:

\begin{itemize}
    \item Flight Position Notice (fl): First on-board estimate of the nature of the transient arriving at with a latency of t$\rm_{fl}$ with a localization radius R$\rm _{fl}$.
    \item Ground Position Notice (g): Refined ground-based position announced at time t$\rm_{g}$ with a localization radius of R$\rm_{g}$. 
    \item Final Position Notice (fi): Updated localization with radius of R$\rm_{fi}$ arriving with a latency of t$\rm_{fi}$. 
    A full HEALPix sky map is released hours after the trigger, enabling credible regions at 50\% and 90\%. If the GRB is detected with instrument (such as \textit{Swift}/BAT) that localizes the transient within arc-minute localization, the final position is not announced. 
\end{itemize}

Multiple flight/ground notices are communicated for a single burst through GCN notices. However, for this study, we consider only the first ones. For all notices (fl, g, and fi), we extract the center of the localization (RA, Dec) and 1$\sigma$ error radius. Since GBM reports circular error regions, we model them as circular Gaussian patches on the sphere. To build a realistic observation scenario, our mock catalog exploits the localization information of real GRBs extracted from the GCN archive\footnote{\url{https://gcn.gsfc.nasa.gov/gcn/fermi_grbs.html}}. 

In order to focus our analysis on poorly localized Fermi/GBM events, we exclude the GRBs detected with Swift/BAT ($\sim$35–40 GRBs/year) which are also seen by Fermi/GBM. This represents around $\sim$15\% of the GRBs seen by GBM but localized jointly with BAT. The target GRBs for this study remains 85\% of the total GRBs seen by GBM.
A more detailed statistical description of the trigger quantities of the arrival time of the notices and the error-region estimates are given in appendix Sect. \ref{secA:trigger_properties}. 

\begin{figure}[ht]
    \centering
    \includegraphics[width=0.8\columnwidth, height = 9cm]{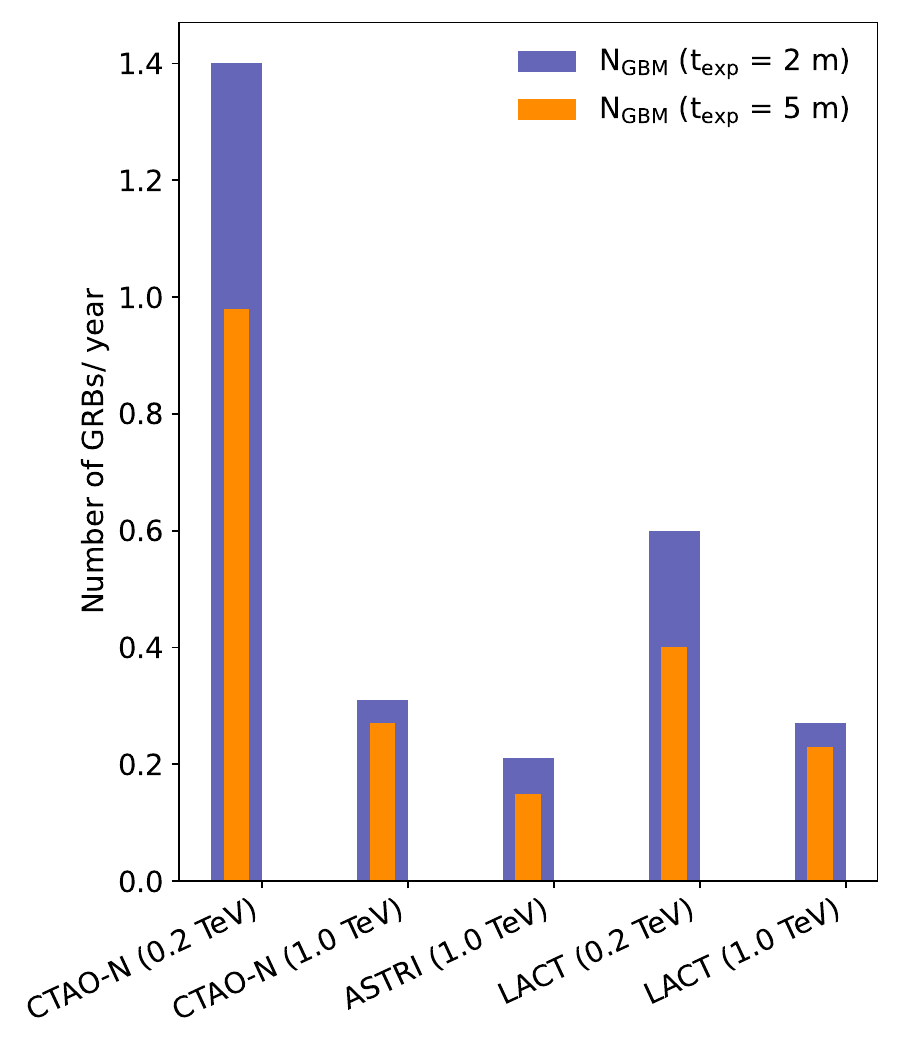}
    \caption{Number of detectable GRBs by CTAO-N, ASTRI and LACT for energies 200 GeV and 1 TeV and for two exposure times of 2 mins and 5 minutes.} 
    \label{fig:VHE_DetGRBs}
\end{figure}

\subsection{Estimates of number of GRBs discovered per year using differential sensitivity at single energies}

To estimate the number of detectable GRBs, we make use of the fraction of GRBs discoverable per year depending on the latency and the exposure time presented in Figs. \ref{fig:LACT_200GeV_dl_exp} and \ref{fig:1TeV_dl_exp}, to predict the number of observable GRBs per year with a fixed exposure of 2 and 5 minutes, given the duty cycle of each detector presented in Sect.~\ref{sec:Detectors}. For the ground based facilities, we assume a duty cycle of about 10\%. 

We make use of the individual light curves of the GRBs as shown in Fig.~\ref{fig:VHE_LCs} for a fixed energy bin E. With two specific exposures of 2 minutes and 5 minutes, we compute the number of pointings that can be executed before the source flux $\rm F^{i}_{E}(t_{obs})$\footnote{Here and throughout the paper, "i" indicates a single GRB.}, where i represents a single GRB, decreases below the limiting flux, $\rm F^{t_{exp}=2m/5m}_{lim} (E)$ (see Eq.\ref{eq:Flim}). In addition, for each GRB, we take into account only the GBM-localization obtained from the ground notice (g) with 1$\sigma$ ($\rm \Delta \Omega^{g, i}_{GBM, 1\sigma}$) and 2$\sigma$ ($\rm \Delta \Omega^{g, i}_{GBM, 2\sigma}$) sky-localization error computed using the error-radius presented in Fig.~\ref{fig:GCN_info_t_R}. 
Regardless of the source brightness, a total of 25 pointings are performed, corresponding to about one hour window after which the flux of the source becomes too weak for ground-based detection (see, e.g., Fig. \ref{fig:LACT_200GeV_dl_exp}). 
For bright GRBs, we compute N$\rm ^{i}_{p}$, the number of pointings during which the flux remains above threshold. Using these two information and the FoV of the instrument ($\Phi^{eff}_{\rm Inst}$), we compute two probabilities: p$\rm ^{i}_{1\sigma}$ and p$\rm ^{i}_{2\sigma}$, defined as: 

\begin{equation}\label{eq:detProb}
     \rm p ^{i}_{1\sigma/ 2\sigma} =  \frac{\Phi^{eff}_{Inst} \times N\rm ^{i}_{p}}{\Delta \Omega^{g, i}_{GBM, 1\sigma/ 2\sigma}}
\end{equation}

If the flux is always below the threshold, a null probability is assigned to the source ($\rm p ^{i}_{1\sigma/ 2\sigma} = 0$). In order to compute the total number of GRBs that can be detected by VHE instruments, we sum the probabilities in Eq.~\ref{eq:detProb} over all the GRBs in the sample. These values are further reduced by the instrument duty cycle ($\mathcal{D}\sim$0.1) and by the probability that the true source location falls within the 1 $\sigma$ ($\mathcal{C}^{\rm {2D}} _{1\sigma} \sim 0.39$) or 2 $\sigma$ ($\mathcal{C}^{\rm {2D}} _{2\sigma}\sim0.89$) GBM localization region. Hence the total number of GRBs that can be potentially detected through the proposed method is about $\rm N^{d, VHE}_{GBM, 1\sigma/2\sigma}$ depending on the strategy of covering $1\sigma$ or $2\sigma$ error-region as stated below in Eq.~\ref{eq:Ntot_VHE}.  

\begin{equation}\label{eq:Ntot_VHE}
    \rm N^{d, VHE}_{GBM, 1\sigma/2\sigma}  = \sum^{N_{GBM}}_{i=1}\rm p ^{i}_{1\sigma/ 2\sigma} \times \mathcal{D}_{Inst} \times \mathcal{C}^{2D}_{1\sigma/ 2\sigma}
\end{equation}

We chose three detectors: CTAO-N (200 GeV and 1 TeV), ASTRI (1 TeV) and LACT (200 GeV and 1 TeV) and their corresponding effective\footnote{In this calculation, we considered half of the FoV of the instrument as the effective FoV. Since the sensitivity of the detectors at the edge of the IACT-camera is significantly reduced, we considered a narrow FoV to have more realistic estimates of detection.} FoV ($\Phi^{eff}_{\rm Inst}$) from Tab.~\ref{tab:Detectors}. With a total number of \textit{Fermi}/GBM-triggered GRBs that will potentially be followed, N$\rm^{GBM}\sim 220$, we compute the number of GRBs with discovered VHE counterparts as shown in Eq.~\ref{eq:Ntot_VHE}. 

\begin{table}
    \centering
    \begin{tabular}{ccccccc}
        \hline
        \multirow{2}{*}{Inst.}    & $\rm \Phi^{eff}_{Inst}$ & \multirow{2}{*}{f$^{\rm d, VHE, max}_{100\ \mathrm{s}}$}  & \multirow{2}{*}{N$\rm ^{GBM}_{2m}$}  & \multirow{2}{*}{N$\rm ^{GBM}_{5m}$}\\
                         & [deg$^2$]  &    &  &  \\
        \hline
        ASTRI  & 25 & 0.02 &  0.20 & 0.15 \\ \hline
        LACT  & \multirow{2}{*}{20} & \multirow{2}{*}{0.13} &  \multirow{2}{*}{0.60} & \multirow{2}{*}{0.40}   \\
        (0.2\,TeV)  &  & &   & &  & \\ \hline
        LACT    & \multirow{2}{*}{20} & \multirow{2}{*}{0.05} &  \multirow{2}{*}{0.27} & \multirow{2}{*}{0.23} \\
        (1.0\,TeV)  &  & &   & \\
        \hline
        CTAO-N   & \multirow{2}{*}{15} & \multirow{2}{*}{0.28}  &  \multirow{2}{*}{1.40} & \multirow{2}{*}{0.98} \\
        (0.2\,TeV)  &  & &   & &  & \\ \hline
        CTAO-N  & \multirow{2}{*}{15} & \multirow{2}{*}{0.06}  &  \multirow{2}{*}{0.31} & \multirow{2}{*}{0.27}  \\
        (1.0\,TeV)  &  & &   & &  & \\ \hline
    \end{tabular}
    \caption{Yearly detection rates of GRBs in the very-high-energy band for different IACT configurations. The effective field of view \(\Phi^{\rm eff}_{\rm Inst}\), the maximum detectable VHE fraction at 100\,s \(f^{\rm d,VHE,max}_{100\,\mathrm{s}}\) and the number of GRBs detected following the procedure proposed in this work with exposures of 2\,min and 5\,min (N$^{\rm GBM}_{2\,\mathrm{m}}$, \(N^{\rm GBM}_{5\,\mathrm{m}}\)).}\label{tab:VHEdet}
\end{table}

According to the procedure mentioned above, we calculated the total number of GRBs detected by ASTRI, LACT, CTAO-N and  based on the tiling strategy. 
The tiling strategy alone provides a detection of about 1 GRBs/year considering 2 minutes or 5 minutes exposure for LACT and CTAO-N for an energy band of 200 GeV. 
The detections at energy band of 1 TeV for ASTRI, LACT and CTAO-N are about 0.3/year. 

Furthermore, assuming a detection rate of 80 GRBs per year with a Swift/BAT-like instrument that delivers well-localized events with $\mathcal{D}=0.1$, the number of GRBs that can potentially be observed is about 8 per year. In addition, the fraction of GRBs that ASTRI can detect at 1 TeV is approximately 0.02 (see Fig. \ref{fig:1TeV_dl_exp}), given a shorter observational latency of less than 2 minutes. Combining these factors, ASTRI is expected to detect about 0.16 GRBs per year at 1 TeV from localized triggers alone, which is {slightly lower than} the rate predicted in this work. For LACT, the fraction of GRBs that are detectable at 200 GeV is about 0.1 (see Fig. \ref{fig:LACT_200GeV_dl_exp}). With the same duty cycle ($\mathcal{D}=0.1$), the number of localized GRBs that can be discovered is about 0.8 per year. This number is comparable with the detection rate of LACT that can be achieved following the method proposed in this study. In case of CTAO-N, we do not provide the number of possible detections for well-localized GRBs, which is the subject of a CTAO Collaboration publication currently in preparation that takes into account the complex responses of the different sub-systems: Small-Sized, Medium-Sized, and Large-Sized Telescopes.

\begin{figure}[h]
    \centering
        \includegraphics[width=\linewidth]{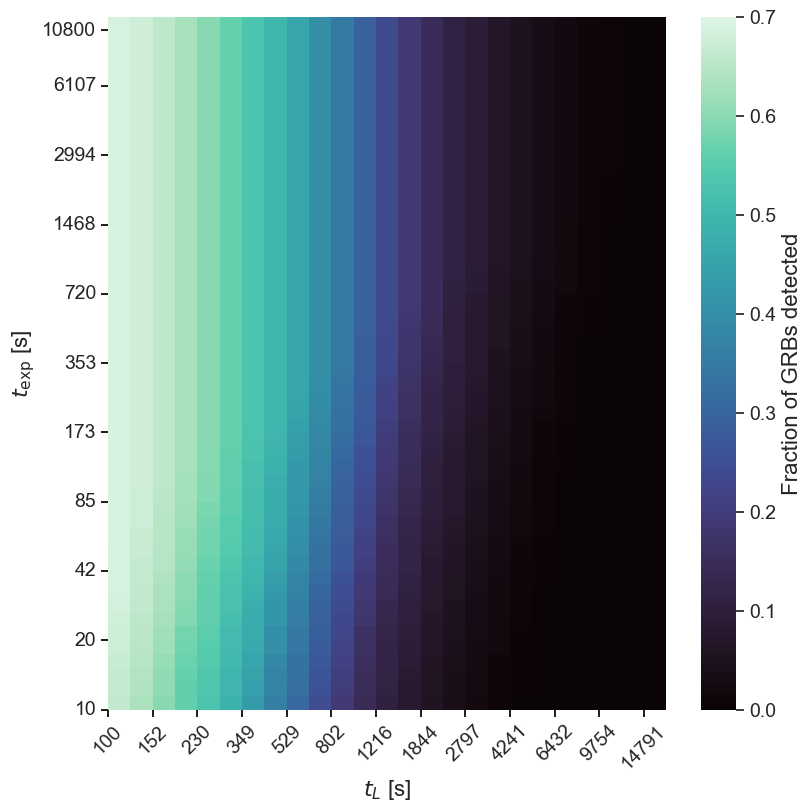}

    \caption{Detectability heatmaps for CTAO-N as a function of latency $t_L$ and exposure time $t_\text{exp}$. The color scale represents the fraction of GRBs in the sample which are detectable at $5\sigma$ by each observatory given their time-dependent spectra. Each of the 220 events from the mock catalog is processed with IRFs for three different zenith angles (20, 40, and 60~deg), meaning that each cell of the heatmap is a fraction over 660 simulated observations. Lastly, after a cutoff around $t_L \sim 30$~min, nearly the entire population of GRBs is no longer detectable.}
    \label{fig:CTAO-results}
    
\end{figure}

\subsubsection{Using time-dependent spectra}
\label{sec:ctao-sensipy}
Due to the need for multiple different telescope pointings to cover the uncertainty region of poorly-localized GRB events, it is necessary to understand the relationship between detectability and the latency ($t_L$) from the onset of the GRB emission. Given that the afterglow models from Section~\ref{sec:LC_VHE_model} contain time-dependent spectra, we can produce even more robust estimations of this relationship for instruments with publicly-available IRFs. Using the open-source simulation tool \texttt{sensipy} \citep[][based on \citealt{Patricelli:2018zfk}]{Green:2026pff} and the CTAO Alpha Configuration IRFs \citep{cherenkov_telescope_array_observatory_2021_5499840}, we can estimate the observation time required for 5~$\sigma$ detections given various latencies, which take into account the entire spectral energy range of the LC simulations.

More specifically, we map the integration time required for a significant detection (\(5\sigma\)) as a function of the delay since the onset of the GRB. Establishing this relationship enables rapid feasibility assessments based on latency. Beyond individual event analysis, this population study reveals the impact of specific observational factors on detectability and provides the necessary input to optimize the strategies presented in Section \ref{sec:observation-strategy}.

To quantify this, we calculate the detection time, $t_{\text{det}}$, by finding the specific exposure duration, $t_{\text{exp}}$, where the average flux of the source equals the CTAO-N $5\sigma$ sensitivity threshold. Mathematically, we solve for the $t_{\text{exp}}$ that satisfies the equilibrium condition \citep[see][]{Green:2026pff}.


\begin{equation}
    F_{\text{avg}}(t_\text{exp}; t_L) =
    F^{\text{CTAO}}{5\sigma}(t_\text{exp})
\label{eq:sensipy-eq}
\end{equation}


Here, the time-averaged flux over the observation window is defined as:


\begin{equation}
    F_{\text{avg}}(t_\text{exp}; t_L) = \frac{S(t_\text{exp}; t_L)}{t_\text{exp}} ,
\end{equation}


where $S$ represents the fluence of the accumulated intrinsic source:


\begin{equation}
    S(t_\text{exp}; t_L) = \int_{t_L}^{ t_L + t_\text{exp}} \int_{E_\text{min}}^{E_\text{max}} \phi(E, t)\ E \ dE \ dt ; .
\end{equation}


In these expressions, $\phi(E, t)$ denotes the time-dependent spectral model detailed in Section \ref{sec:LC_VHE_model}, integrated over the energy limits $E_\text{min}$ and $E_\text{max}$ corresponding to the IRFs. For the temporal integration limits, we define $t_0 = 0$ as the time of GRB onset.

Figure~\ref{fig:CTAO-results} contains so-called detectability heatmaps, which indicate the fraction of events from the overall sample that are detectable as a function of $t_L$ and $t_\text{exp}$. Importantly, the presence of both LSTs and MSTs allows CTAO-N to detect more events, particularly at latencies shorter than $\sim10$~min. In addition, after latencies around $\sim30$~min, nearly the entire GRB population is no longer detectable by either site.

\begin{figure}[ht]
    \centering
    \includegraphics[width=\columnwidth, height=9cm]{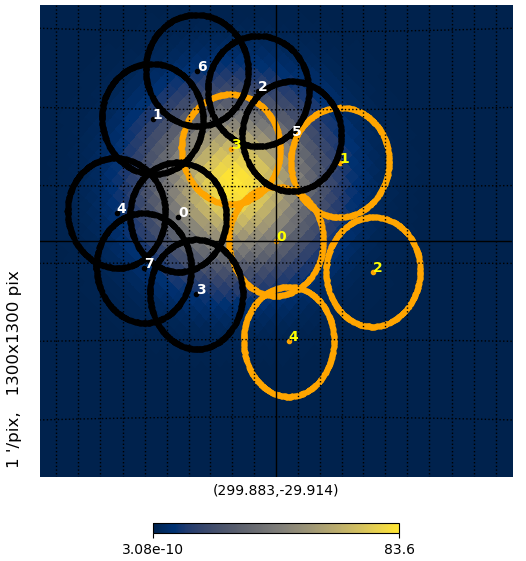}

    \caption{Simulated follow-up observation strategy for the Ground reconstruction (arrival time 22:21, orange tiles) and Final reconstruction (arrival time 22:30, black tiles) of the same Fermi-GBM GRB, GRB 210731931. The GRB Final localization is presented in the background to show the coverage of the GRB Ground-alert strategy, used as input to derived the GRB Final-alert strategy.  Numbers corresponds to the ground (yellow) and final (white) campaign on each localization, which together represent the full observing campaign. The radius of the telescope FoV of the pictures correspond to FoV$_{radius}$ = 2.5$^{\circ}$, emcompassing approximately half of the nominal telescope sensitivity. The contribution from the outer FoV region is omitted for clarity.}
    \label{fig:tilepyresults}
\end{figure}
\begin{table}[ht]
\centering
\begin{tabular}{lcccc}
\toprule
Pointing duration [min] & \multicolumn{2}{c}{2} & \multicolumn{2}{c}{5} \\
\cmidrule(lr){2-3} \cmidrule(lr){4-5}
FoV [deg] & 2.5 & 4 & 2.5 & 4 \\
\midrule
N. follow-ups & 35 & 39 & 31 & 37 \\
N. follow-ups (with cuts) & 7 & 9 & 6 & 9 \\
N. detected & 1 & 4 & 1 & 3 \\
Percent det. [\%] & 0.5 & 2.0 & 0.5 & 1.5 \\
\% det. (with cuts) & 14.3 & 44.4 & 16.7 & 33.3 \\
Avg. time to det. [min] & 18.0 & 15.2 & 45.0 & 25.1 \\
Avg. pre-trial sig. [\(\sigma\)] & 46.2 & 38.8 & 11.2 & 20.5 \\
\bottomrule
\end{tabular}
\caption{Tilepy results for sources with follow-ups. A similar subset of events are followed up for each pointing duration and FoV pair. By applying strict filters on which events are chosen for observation, we can further reduce the number of follow-ups without losing any detectable events, leading to a more efficient campaign. In this case, the selected cuts are zenith angle $<$ \SI{50}{\deg} and latency $t_L <$ \SI{30}{\min}. Generally, shorter pointing durations accelerate source acquisition, enabling earlier detections and enhanced pre-trial significance.}
\label{tab:tilepy-results}
\end{table}

\subsection{Scheduling observations with \texttt{tilepy}}
\label{sec:tilepy}

Simulation of follow-up campaigns were carried out for the CTAO-N site using \texttt{tilepy} \citep{Seglar_Arroyo_2024}. Four observation campaign are designed based on the insights obtained in previous section - we combine two fixed pointing durations of \SI{2} and \SI{5}{\min}, and two array FoV radius of \SI{2.5} and \SI{4}{\deg}. The entire GRB catalog is simulated for each combination of the aforementioned parameters. The choice of pointing durations is motivated by 
Figs. \ref{fig:LACT_200GeV_dl_exp}, \ref{fig:1TeV_dl_exp}, and \ref{fig:CTAO-results}, where we note that the optimal exposure time has to be shorter in order to optimize the detection (see also \citealt{CTAO2026}). 
In addition, the results indicated in Sect.~\ref{sec:ctao-sensipy} indicate that after a cutoff of around \SI{\sim30}{\min}, most GRBs in the sample are no longer detectable by CTAO-N. The FoV radius of \SI{2.5} and \SI{4}{\deg} chosen correspond to the effective FoV of CTAO LSTs and MSTs, respectively.
{An example of the scheduling of the observations is shown in Fig~\ref{fig:tilepyresults}.}

After computing the list of pointings for each event with \texttt{tilepy}, the \texttt{sensipy} package is once again used a posteriori to determine whether each source is detected. In addition, the significance of detection in each successful pointing, as well as the cumulative significance in case observed regions overlap, is calculated. In this simulated year of events regardless of the chosen configuration, between 30-40 follow-ups are carried out, and a maximum of 4 of those events are detected above $5\sigma$. With the \SI{2.5}{\deg} FoV configuration, only a single event is detected, illustrating how the wider FoV of the MSTs play a crucial role in increasing detection rates.

Although we cannot increase the number of events per year, we can increase the efficiency of our follow-up campaigns. By restricting follow-up observations with cuts in observability parameters, we can maximize the ratio of detected events with respect to the total number of follow-up observations. In particular, we found two cuts that optimize this ratio: zenith angle $<$ \SI{50}{\deg} and latency $t_L <$ \SI{30}{\min}. In applying these cuts, we effectively eliminate 75\% of follow-ups that are not likely to result in a detection and optimizing the use of telescope time. 

We also see that the 2-minute pointing campaign led to detections 40-60\% sooner compared to the 5-minute campaigns. In addition, we find that on average the significance of detections doubles in the 2-minute scenario. Due to the rapidly decreasing flux in the simulated spectra, these data show that it is important to cover as much of the uncertainty region as quickly as possible in order to optimize the number of detections.  Lastly, we note the necessity of a real-time analysis (RTA), which can properly understand when signal is seen to stop the tiling and continue exposing the source location. The results of these follow-up campaigns are summarized in Table~\ref{tab:tilepy-results}.

\section{Discussion and Conclusions}

The aim of this study is to assess the feasibility of detecting very-high-energy emission from poorly localized GRBs using a realistic strategy tailored for sources detected by wide field gamma ray instruments such as \textit{Fermi}/GBM. While GRBs are among the most luminous explosions in the Universe, the detection of their VHE counterparts is still challenging. So far, GRBs follow-up observations have been highly biased towards well-localized GRBs \citep{Abe:2025bvx}. Although the observational strategy is straightforward, the number of triggers is a factor of three less than that of the MeV detectors. 
To maximize scientific return, we present an optimized observational strategy based on rapid tiling, designed to benefit from large number of GRBs detected by MeV detectors, which are typically poorly-localized.

We emphasize how ground-based atmospheric Cherenkov telescopes requires complementary observational strategies in GRB follow-up with both well-localized events (\citep{CTAConsortium:2019cbf} and CTAO consortium paper in prep.) (localization better than an arcminute, typically from hard X-ray instruments such as \textit{Swift}/BAT) and poorly localized events (error regions of several square degrees, typically from MeV instruments such as \textit{Fermi}/GBM). The next generation of ground-based IACTs, including ASTRI, LACT and CTAO-N will offer outstanding sensitivity in the 0.02-5 \,TeV range and, thanks to their significantly larger fields of view, will allow the follow-up of not only well-localized GRBs but also those with larger localization uncertainties. Nearly all of these facilities can respond rapidly to external triggers, with typical slew times of about 30\,s to a minute, and provide broad energy coverage. However, the relatively low duty cycles ($\sim$10\%) present a limitation. This highlights the importance of systematically following up the large number of GRBs detected by MeV instruments (e.g., \textit{Fermi}/GBM; factor of three larger than hard-X-ray instruments) to enhance the likelihood of observing GRBs at TeV energies.

To simulate one year of observed GRBs with their corresponding VHE light curves, the key ingredients for building a GRB catalog are the redshift, fluence (and the $E^{\gamma}_{\mathrm{iso}}$), and sky localization. For this reason, we start from the distributions of redshifts ($P_{1}$) and fluences ($P_{2}$), obtained from the Greiner catalog and the GBM catalog, respectively. We build a third population, namely $P_{3}$, of GRBs commonly detected by GBM and \textit{Swift}/XRT in the last 16 years.
From $P_{1}$ and $P_{2}$ we extract random couples ($z$, $S_{\gamma}$) and make sure that the corresponding $E^{\gamma}_{\mathrm{iso}}$ lies above the GBM detectability line. Each simulated GRB is assigned sky-localization information by matching its fluence to the closest event in the real sample $P_{3}$, providing realistic uncertainty regions and follow-up constraints. Overall, this methodology yields a synthetic one-year GRB population that is observationally consistent and suitable for VHE follow-up studies.

Once the simulated population of GRB is processed ($P_{5}$), we simulate the afterglow spectra and compute the VHE light curve for each GRB, taking as input for the simulations the redshift and $E^{\gamma}_{\mathrm{iso}}$. For the afterglow modeling, we use the set of microphysical parameters which was shown to best represent the observed correlations in X-ray and GeV bands. The VHE fluxes computed in this work should be interpreted as conservative lower limits. We model only the afterglow emission using a synchrotron self-Compton framework, neglecting additional physical processes which might enhance the TeV output. Moreover, we do not include any contribution from the prompt emission, which may yield bright and short VHE flashes. Because prompt emission is highly uncertain and variable, excluding it allows us to be conservative, but it also means that our detection rates represent a lower boundary of what might be achievable for real GRBs. 
Finally, we include attenuation by EBL, which strongly suppresses TeV emission at redshifts $z > {0.3}$. This effect alone already reduces the number of detectable VHE events significantly.

Using the clustering properties of X-ray afterglows, normalizing the flux by the fluence, we investigated whether the observed correlation between X-ray and VHE light curves is reproduced from the model used to simulate the afterglow. Our results show that the correlation is reproduced at early times, while it is less evident at later times. Since the afterglow of the GRBs is rapidly declining with a temporal index lower than $-1$ and the sensitivity of the IACTs vary as t$^{-0.5}$, a small exposure increases the chances of discovering the VHE counterpart. 

Our simulated GRB catalog contains the localization accuracy and alert delays of GBM-triggered GRBs that are relevant for ground-based follow-up. For each burst, it compiles, along with  arcminute localizations from \textit{Swift}/XRT, the full sequence of notices from GBM (flight, ground, and final), including their associated sky regions and arrival times. This catalog enables a statistical evaluation of how rapidly and how significantly the localization improves over time. The cumulative distributions of these parameters indicate that, for 68\% of events, the flight notice is issued within $30$~s with a typical error radius smaller than $20^\circ$, the ground notice is sent after about $50$~s with error radii below $10^\circ$, and the final notice becomes available within $600$~s with an error radius under $7^\circ$. Over the same percentile interval, the localization centroids move on average by roughly $13^\circ$ between the flight and ground notices, and by about $6^\circ$ on average between the ground and final notices, thereby quantifying how the probability region both contracts and shifts with time. 
This information is crucial for developing an observational strategy to repoint IACTs in response to real-time alerts, while accounting for the additional delay caused by telescope slewing. Because only about 15\% of GBM-detected GRBs are also observed simultaneously by \textit{Swift}/BAT, a final notice is often not issued.

The expected GRB detection rates for ground-based VHE observatories are highly sensitive to the chosen observing strategy, in particular, on the latency time of observation and the exposure. Our results show that if observations begin with delays of about 10–15 minutes ($10^{3}$\,s), only a small subset of GRBs remains observable at VHE energies. To obtain realistic estimates of the detection per year, we therefore adopt a representative latency of roughly 1000~s and evaluate detectability using individual GRB light curves in fixed energy bins drawn from our mock population. For each burst, we compute how many telescope pointings with a fixed two-minute or five minutes exposures can be carried out before the source flux drops below the sensitivity limit of the instrument. We restrict our analysis to ground-based alerts (g) that are received within one minute of the transient trigger. As the sky localization regions reported in these ground alerts are typically large, multiple pointings are needed to tile the corresponding area. We examine two tiling schemes based on the $1\sigma$ and $2\sigma$ GBM error circles, and for each GRB we calculate the detection probability as the ratio between its localization area and the total field of view covered by the sequence of brightness-limited pointings. Summing these probabilities over the entire GRB sample, and incorporating both the $\sim 10\%$ duty cycle of ground-based facilities and the geometric probability that the true source position lies within the assumed Gaussian error contour, provides an estimate of the number of GRBs detected per year. This approach shows that targeting poorly localized GRBs can yield about 0.5-2 GRBs per year which further complements strategies focused on well-localized events. 

We further substantiate the results of our sensitivity-based analysis by performing a more detailed study that uses the instrument response function and focuses exclusively on CTAO-N as a representative example. In the latter analysis, we confined ourselves to a significantly smaller field of view (the effective FoV), because the sensitivity decreases toward the edge of the camera relative to its center. Our results indicate that, given the uncertainty in the sky localization provided by GBM, between 1 and 4 GRBs per year may be detected. These values are slightly higher than those obtained from the probabilistic approach based solely on differential sensitivity, since the true burst position can be covered with more than one tile while performing a tiling inside the sky-region. We also find that imposing an effective zenith-angle limit of 50$^{\circ}$ and extending the tiling to times later than about one hour after the trigger do not increase the number of detections and can therefore be excluded to reduce the total observation time. A shorter exposure of 2 minutes is preferred to longer exposures of 5 minutes, as expected, because the transient flux decays faster than t$^{-1}$. An annual observing budget of about 20 hours,  according to the strategy presented here, can result in 1–4 GRB detections per year. Moreover, incorporating CTAO-S could approximately double the yearly detection rate. 

{For ASTRI, we find that the TeV detection rate is similar to CTAO-N and LACT while having worse sensitivity than the others. This behavior arises due to the large FoV that exceeds 20 deg$^2$. Additionally, the follow-up of poorly localized GRBs (see Fig. \ref{fig:VHE_DetGRBs}) contributes substantially to the total number of detections, mainly because tracking a larger number of triggers from MeV instruments increases the probability of a successful detection.}

Although this work focuses on GRBs observed with \textit{Fermi}/GBM, the results are not restricted to this particular instrument. The observing strategies and conclusions we present can be directly applied to future MeV missions. The GBM localization folds the complex, irregular sky-localization pattern (arising from several MeV detectors pointing in different directions) into an ellipse, which can therefore fail to include the true source position. 
For comparison, the probability that the true XRT position lies within the 1$\sigma$ GBM localization region is only about 30\% (rather than $\sim$40\%), and about 70\% (instead of 90\%) for the 2$\sigma$ region (see also \citealt{Connaughton:2014xha}). Current and future missions such as Crystal Eye \citep[CE; ][]{2026APh...17403171A, 2023NIMPA104968045B} and GRINTA \citep{Rodi:2025cnh} are expected to detect even more GRBs per year than GBM, likely with enhanced localization performance. Furthermore, the sky-localization regions provided by GRINTA and CE are expected to be approximately circular and may offer tighter constraints on the true source position. In this setting, optimized tiling strategies for ground-based VHE follow-up will become increasingly crucial. The methodology developed here offers a general framework for maximizing the chances of detecting TeV afterglow emission from the large GRB samples that will be discovered by next-generation MeV instruments. Consequently, with future MeV detectors providing more precise localizations and future IACT facilities, the prospects for detecting TeV counterparts of poorly localized GRBs are very encouraging.

\textit{Software:} This work made use of
\textsc{Gammapy}~\citep[v1.3;][]{gammapy:2023};
\textsc{tilepy}~\citep{Seglar_Arroyo_2024}; and
\textsc{sensipy}~\citep{Green:2026pff}.

\begin{acknowledgements}
The authors thank Marina Manganaro, Qi Feng, Barbara Patricelli, Iftack Sadeh, Francesco Longo, Giancarlo Ghirlanda and Lara Nava for insightful comments and suggestions. BB acknowledges financial support from the Italian Ministry of University and Research (MUR) for the PRIN grant METE under contract no. 2020KB33TP. G.O., A.I. and P.T. acknowledge support from the ASI-GSSI contract n. 2025-5-U.0: “Gamma-ray bursts: a probe of multi-messenger and extreme Astrophysics”. MB, GO, BB and S.Mo acknowledge the ACME project, which has received funding from the European Union’s Horizon Europe Research and Innovation program under Grant Agreement No. 101131928. FS acknowledges ANR (French National Research Agency) for its support of the project "Multi-messenger Observations of the Transient Sky (MOTS)" no. ANR-22-CE31-0012. The authors acknowledge the Fermi and Swift teams for making their data publicly available. 

\end{acknowledgements}

\bibliographystyle{aa}
\bibliography{references}  



\appendix 

\section{X-ray Light Curves Analysis} \label{sec:appendix}

This section gives a detailed description of the analysis of the X-ray light curves used in this work. The results are summarized in the main text (Sec.~\ref{sec:xrt}). 

The analysis is based on the GRB populations $P_{3}$ and $P_{4}$ defined in Table~\ref{tab:Catalog}, corresponding to GRBs jointly detected by \textit{Fermi}/GBM and \textit{Swift}/XRT, and the subset of these events with a measured redshift. For each GRB, we analyze the temporal evolution of the X-ray flux observed by \textit{Swift}/XRT in the energy band 0.3-10 keV. 

X-ray light curves are known to display complex temporal behavior, often deviating from a simple power-law decay. Many GRBs show early-time plateaus, flares or re-brightening episodes. To account for this complexity, we modeled each XRT light curve using a piecewise power-law representation implemented with the \texttt{pwlf} Python package. 

Our analysis is primarily focused on the standard afterglow decay, therefore the flaring activity is filtered from the light curves. In order to do this, we adopt a strategy inspired by previous studies \citep{2009MNRAS.397.1177E,2010A&A...519A.102E}. Significant flares are identified by searching for consecutive rising bins with a total flux increase of at least $2\sigma$. For each flare, the peak is defined as the maximum flux prior to the decline, and the end of the flare is identified through a change in the slope of the light curve. After filtering these intervals, the piecewise fit is performed such that once the first negative temporal slope is reached, marking the onset of the afterglow decay, all subsequent slopes are required to remain negative. 
This approach is physically motivated by the expected monotonic decay of the afterglow. Plateaus and flares, although not fully understood, occur in only a fraction of GRBs and are treated in a way that preserves early-time features potentially relevant for very-high-energy studies. Consequently, the fit quality may be reduced in cases where flares remain in the data, but the overall decay behavior is still captured. 

Following the methodology of \citet{2012MNRAS.425..506D}, we normalized the light curves by the fluence (Flux/Fluence) and interpolated all LCs onto a common time grid. This allowed us to compute the median flux evolution and the corresponding $1\sigma$ and $2\sigma$ uncertainty regions for both the full sample and the subset of GRBs with measured redshift. This procedure enables a statistically robust characterization of the typical X-ray afterglow behavior while accounting for variations between individual bursts.  

To assess whether the selected sample is representative of the broader GRB population, we compared its redshift and fluence distributions with those of Greiner and \textit{Fermi}/GBM catalogs, respectively. The mean redshift in Greiner’s catalog is ⟨z⟩=1.99, while our sample yields a comparable mean of ⟨z⟩=1.91 (see Fig.\ref{fig:GRB_catagorization}). This close agreement indicates that our redshift-selected subsample does not strongly deviate from the general observed GRB population. 

\begin{figure}
    \centering
    \includegraphics[width=\columnwidth]{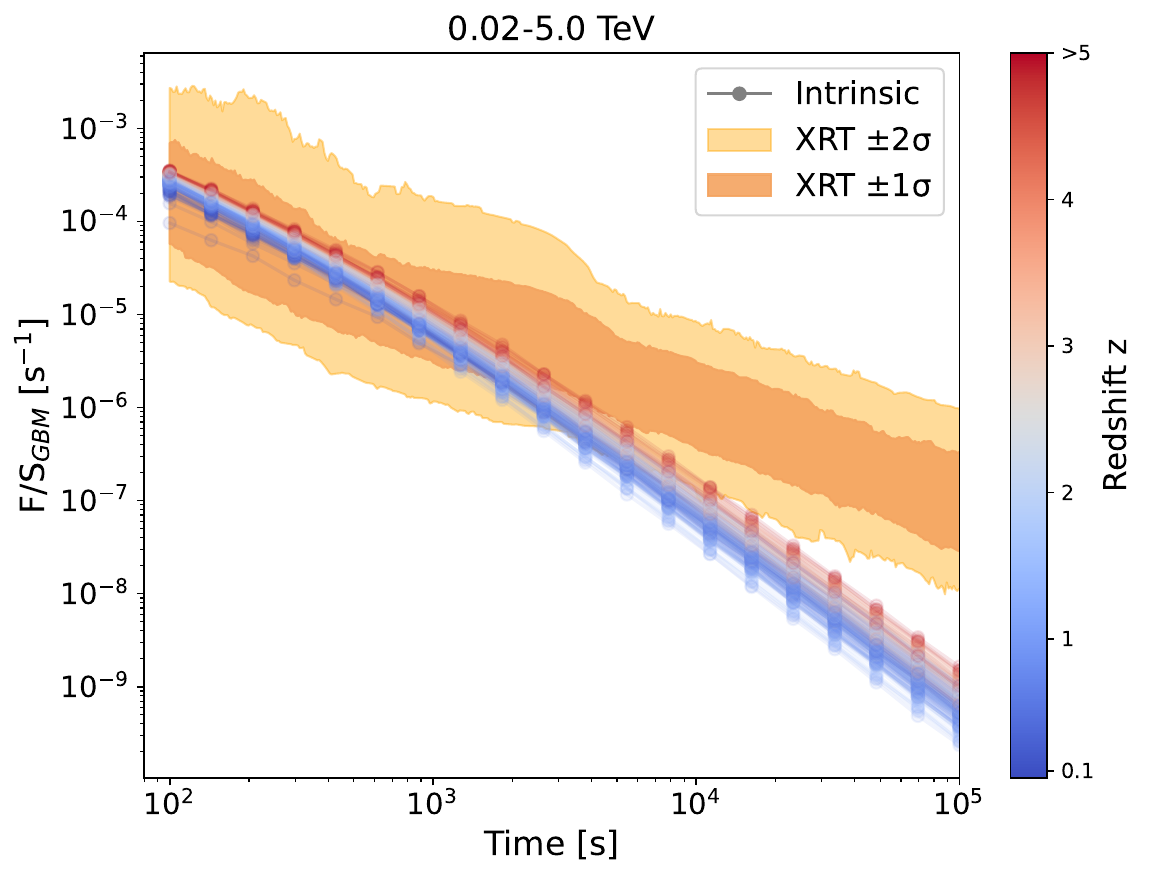}
    \caption{The intrinsic VHE light curves of simulated 220 long GRBs in the energy range of 20 GeV--5 TeV. The shaded region indicates the XRT clustering.}
    \label{fig:XRT_VHEflux}
\end{figure}

For the fluence, the distribution of our catalog mirrors that of the full \textit{Fermi}/GBM dataset. In particular, the \textit{Fermi}/GBM catalog peaks at $\sim 3 \times 10^{-6}\,\mathrm{erg\,cm^{-2}}$, which is essentially the same peak value observed in our sample (see Fig. \ref{fig:XRT-GRB_catagorization}). This similarity further supports that the joint catalog represents the same underlying population of GRBs.
Fig.~\ref{fig:XRT_VHEflux} shows the temporal evolution of the fluence normalized X-ray (\textit{Swift}/XRT) and simulated intrinsic VHE light curves, which display a similar clustering behavior and a strong correlation in normalization and temporal evolution during the early afterglow phase.
 
Taken together, these comparisons suggest that the $P_{3}$ sample used in this work is well defined and broadly representative, rather than being strongly biased toward a particular subset of bursts. This makes it a robust dataset for population studies relevant to VHE follow-up observations.

\begin{table}[h!]
\caption{Scatter of $F_{XRT}/S_{GBM}$ at selected times. {Note: This should be estimated as a Gaussian sigma/ dex.}}\label{tab:XCorrScatter}
\begin{tabular}{lcccc}
\hline
 & t$=10^{3}$ s & t$=10^{4}$ s & t$=10^{5}$ s & \\
\hline
\hline
\multirow{2}{*}{$F_{XRT}/S_{GBM}$} 
 & $2.77\times10^{-5}$ & $2.81\times10^{-6}$ & $3.04\times10^{-7}$  & $1\sigma$\\[-2pt]
 & $1.82\times10^{-4}$ & $8.41\times10^{-6}$ & $9.60\times10^{-7}$ & $2\sigma$\\[-2pt]
\hline
\multirow{2}{*}{$L_{XRT}/E_{iso}$} 
 & $4.44\times10^{-5}$ & $3.47\times10^{-6}$ & $2.76\times10^{-7}$  & $1\sigma$\\[-2pt]
 & $2.92\times10^{-4}$ & $1.12\times10^{-5}$ & $6.66\times10^{-7}$ & $2\sigma$\\[-2pt]

\hline
\end{tabular}
\end{table}

\begin{figure}[h]
    \centering
    \includegraphics[width=\columnwidth]{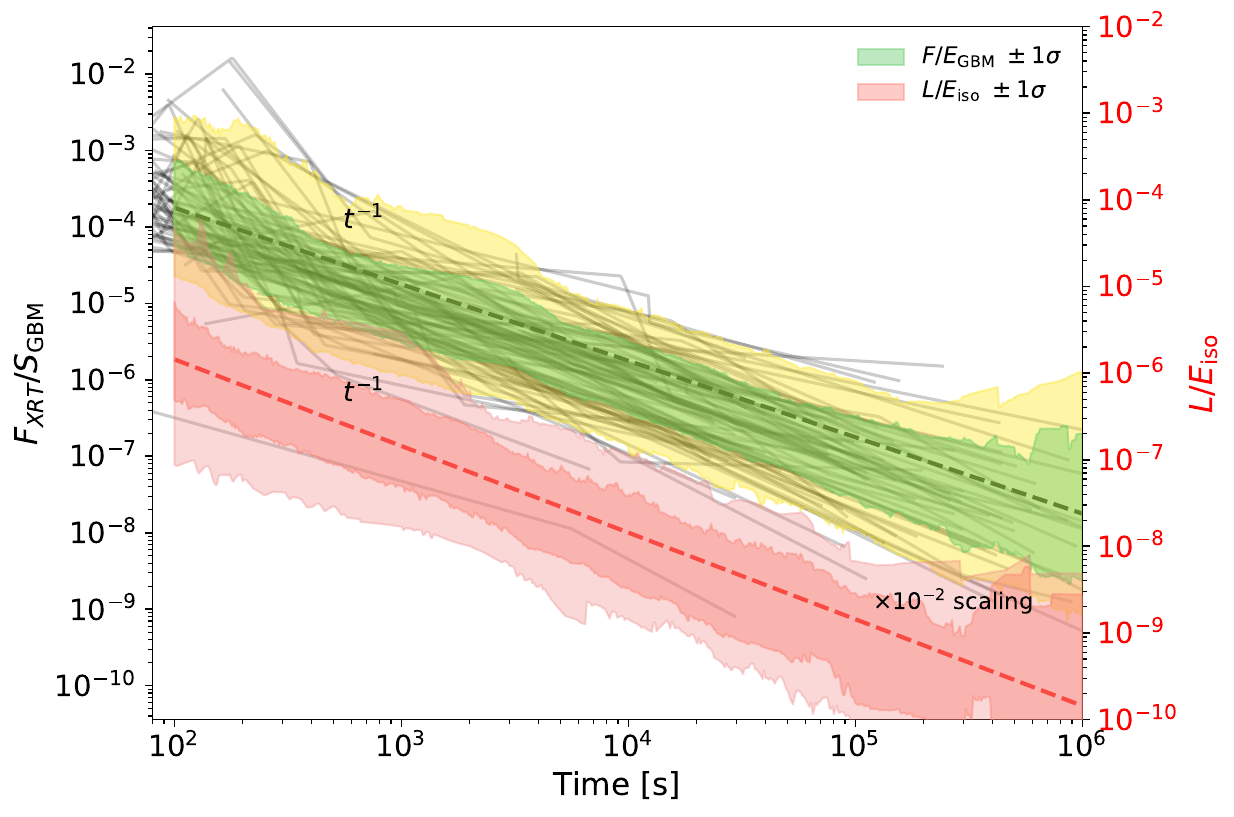}
    \caption{XRT light curve clustering in the energy band of 0.3-10 keV. The shaded region indicate 1$\sigma$ and 2$\sigma$  clustering of XRT light curves, as in \citet{2012MNRAS.425..506D}, in the energy band of 2-10 keV.}
    \label{fig:XRT_clustering}
\end{figure}

\begin{figure}[h]
    \centering
    \includegraphics[width=\columnwidth, height=7cm]{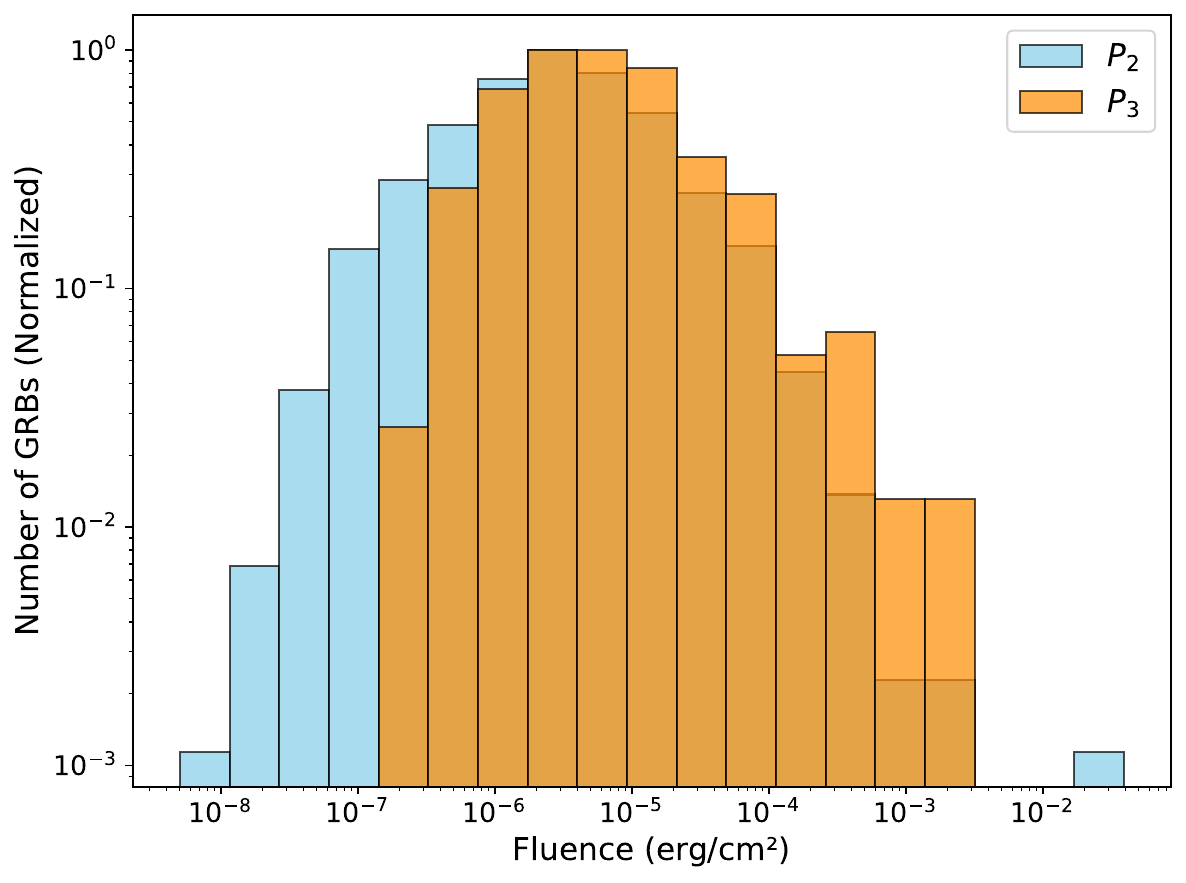}
    \caption{Comparison of the fluence distribution of the populations $P_{2}$ (light blue) and $P_{3}$ (orange). The XRT-GBM sample is moderately biased toward higher fluences, reflecting the higher porbability of successful X-ray follow-up for brightter GRBs, while still tracing the same underlying GRB population.}
    \label{fig:XRT-GRB_catagorization}
\end{figure}

\section{Comparison of SSC microphysical parameters}

In this section, we summarize and compare the SSC microphysical parameters inferred for GRBs with detected VHE emission in the literature, and the benchmark parameter set adopted in this work. Table~\ref{tab:ssc_params_comparison} collects representative SSC modeling results for several TeV-detected GRBs, including GRB~190114C, GRB~190829A, GRB~201216C, and GRB~221009A, as reported by different studies using broadband afterglow observations. The table highlights the range of electron energy distribution indices and equipartition parameters ($\epsilon_e$, $\epsilon_B$) inferred from observations, together with the inferred density of the external medium.

\begin{table*}[t]
\centering
\small
\setlength{\tabcolsep}{6pt}
\renewcommand{\arraystretch}{1.25}

\begin{tabular}{l l c c c c p{2.5cm}}
\hline
GRB & Reference & Medium & $p$ & $\epsilon_e$ & $\epsilon_B$ & $A_{*}$ \\
\hline

190114C &
\citet{MAGIC:2019irs} &
ISM &
2.6 &
$7\times10^{-2}$ &
$8\times10^{-5}$ &
0.5
\\

190114C &
\citet{Aguilar-Ruiz:2025ect} &
Wind-like &
$2.63^{+0.05}_{-0.05}$ &
$4.73^{+0.94}_{-1.22}\times10^{-2}$ &
$3.65^{+0.18}_{-0.33}\times10^{-4}$ &
$5.52^{+1.25}_{-0.82}\times10^{-3}$
\\ 

190114C &
\citet{DP2021} &
 Wind $\&$ ISM &
$2.5$ &
$\approx 0.1$ &
$\approx 6 \times 10^{-3} \star$ &
$\approx 0.1$
\\ 

190829A &
\citet{2022ApJ...931L..19S} &
Homogeneous &
2.01 &
$3^{+2.9}_{-1.7}\times10^{-2}$ &
$2.5^{+3.5}_{-1.3}\times10^{-5}$ &
\\

201216C &
\citet{Abe:2023nhj} &
Wind &
$2.1^{+0.5}_{-0.05}$ &
$0.08^{+0.82}_{-0.07}$ &
$2.5\times10^{-3}$ &
$2.5\times10^{-2}$
\\

221009A &
\citet{Banerjee:2024hxp} &
ISM &
$2.7^{+0.1}_{-0.2}$ &
$\lesssim1.3\times10^{-2}$ &
$\lesssim1.6\times10^{-5}$ &
 \\

221009A &
\citet{Khangulyan:2023srq} &
wind seems preferable &
  &
 &
$\sim3\times10^{-3}$ &
 \\

GRB population &
\citet{Tiwari:2025tgr} &
Wind &
2.3 &
0.1 &
$10^{-4}$ &
0.1
\\

\hline
\end{tabular}

\caption{Comparison of synchrotron self-Compton (SSC) microphysical parameters reported for TeV-detected GRBs and the benchmark parameter set adopted in this work.
$^\star \epsilon_B$ reported is at 90 s (since the GRB trigger time) and it decreases by a factor of 2 in the following temporal bin at 145 s. }
\label{tab:ssc_params_comparison}
\end{table*}

\begin{figure}
\centering
\includegraphics[width=\columnwidth]{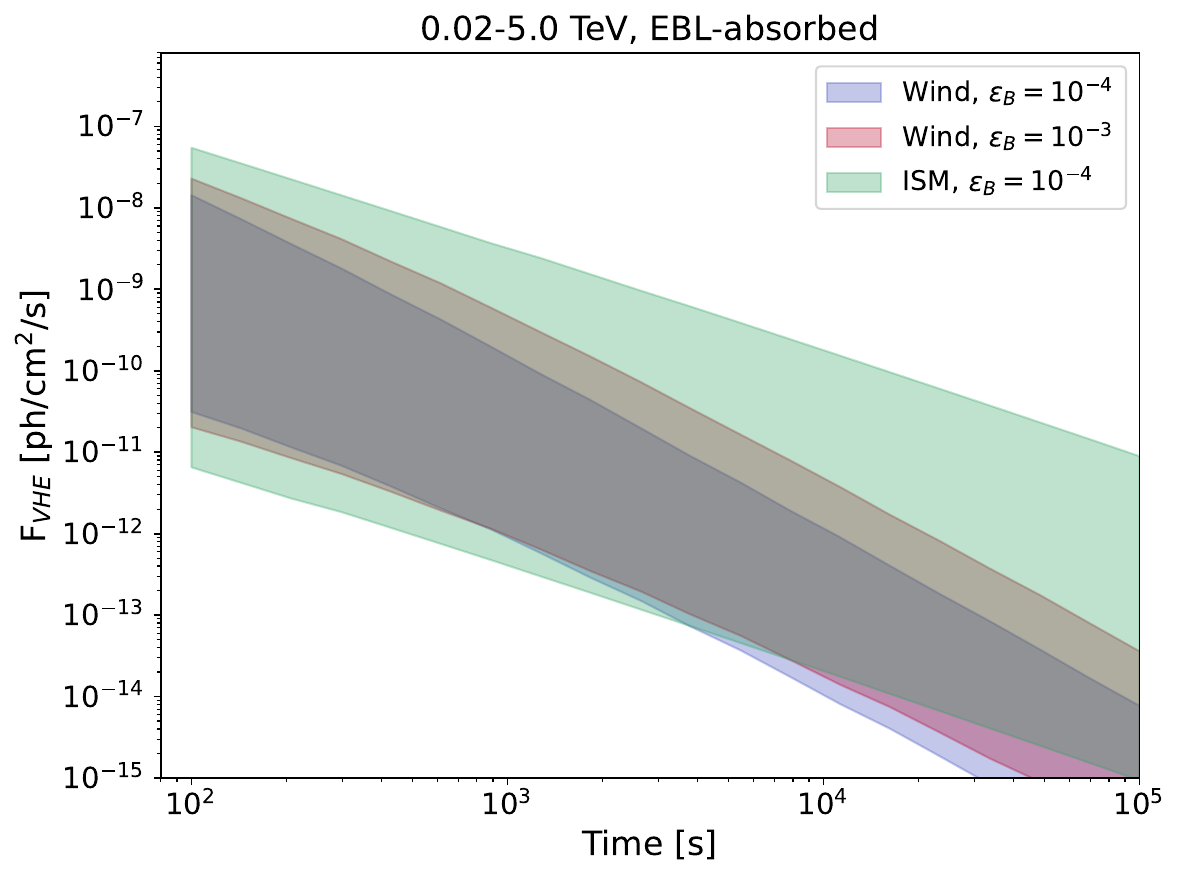}
\caption{Simulated VHE light curves (0.02-5\,TeV, EBL-absorbed) for the population $P_5$, for a wind-like medium with $\epsilon_B=10^{-4}$ (blue), a wind-like medium with $\epsilon_B=10^{-3}$ (red), and a homogeneous (ISM) medium with $\epsilon_B=10^{-4}$ (green).}
\label{fig:lc_medium_comparison}
\end{figure}

To assess how sensitive our results are to the choice of parameters, we recomputed the VHE light curves for the full simulated population ($P_5$) under three configurations: our fiducial wind-like medium with $\epsilon_B=10^{-4}$, a wind-like medium with $\epsilon_B=10^{-3}$, and a homogeneous (ISM) medium with $\epsilon_B=10^{-4}$, keeping all other parameters fixed. We stress that this benchmark set is not intended as a claim on the true nature of the circumburst medium, but as a representative choice falling within the range of values reported in the literature, given the degeneracy among these microphysical parameters. Fig.~\ref{fig:lc_medium_comparison} shows the resulting envelope of light curves (0.02-5\,TeV, EBL-absorbed) for the three cases: the predicted fluxes are remarkably similar across configurations, particularly at early times, with no substantial impact on our conclusions.

\section{Comparison of population of GRBs in this work and intrinsic GRB population}

\begin{figure}[ht]
    \centering
    \includegraphics[width=\columnwidth]{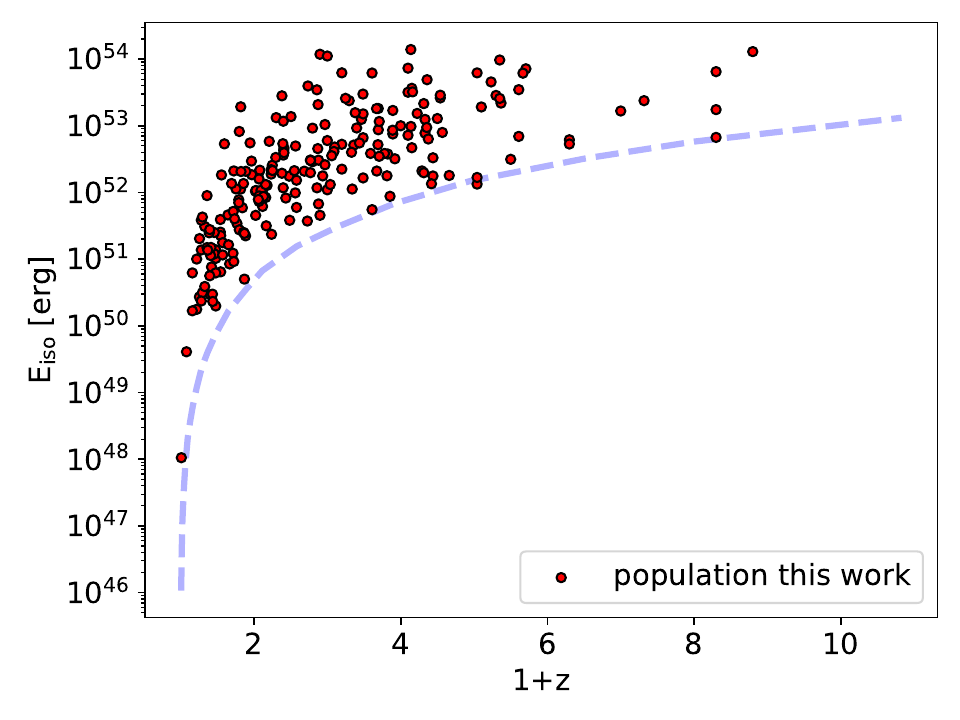}
    \caption{The simulated 220 long GRBs (red points) and the sensitivity limit of \textit{Fermi}/GBM (blue broken line) from \citet{2023ApJ...952L..42L}.} 
    \label{fig:detection_limit}
\end{figure}

\begin{figure*}[h]
    \centering
    \includegraphics[width=\columnwidth, height=7cm]{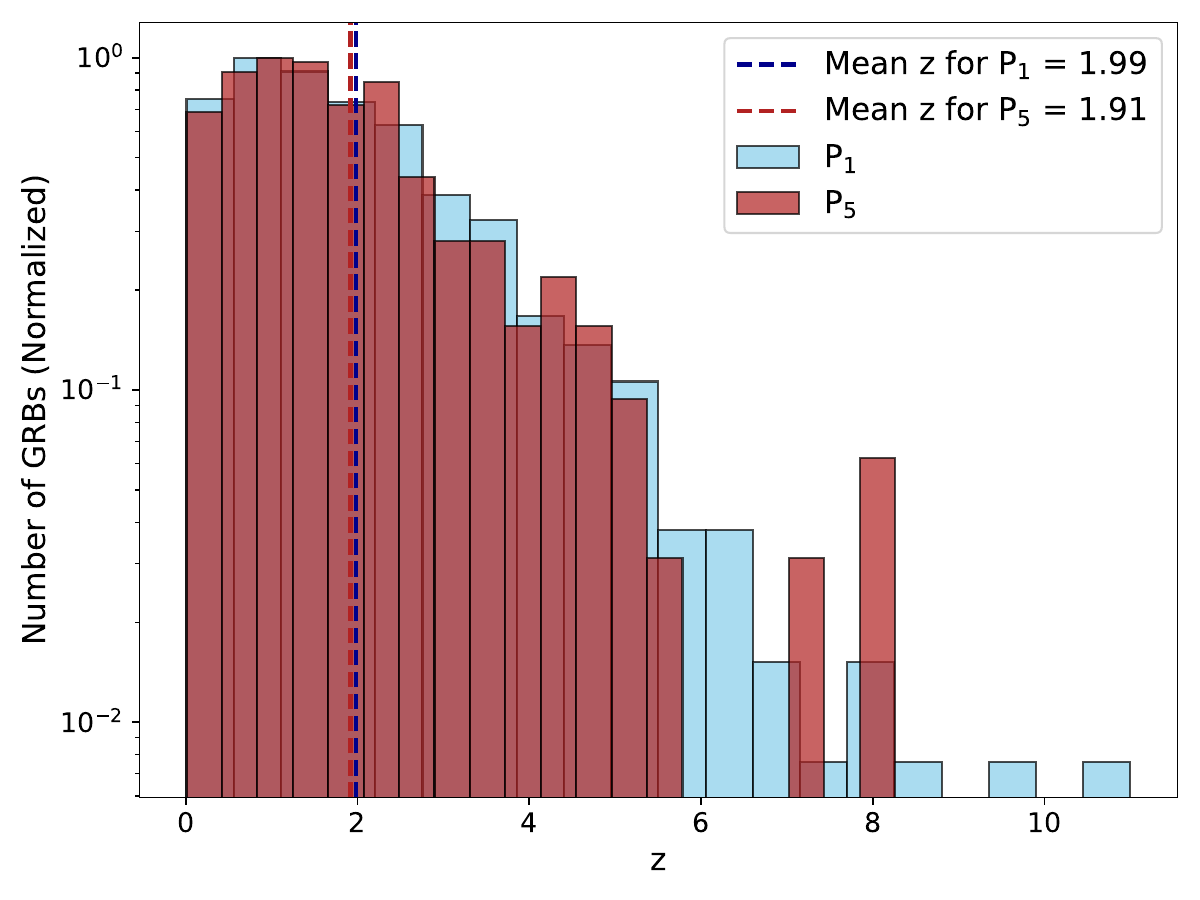}
    \includegraphics[width=\columnwidth, height=7cm]{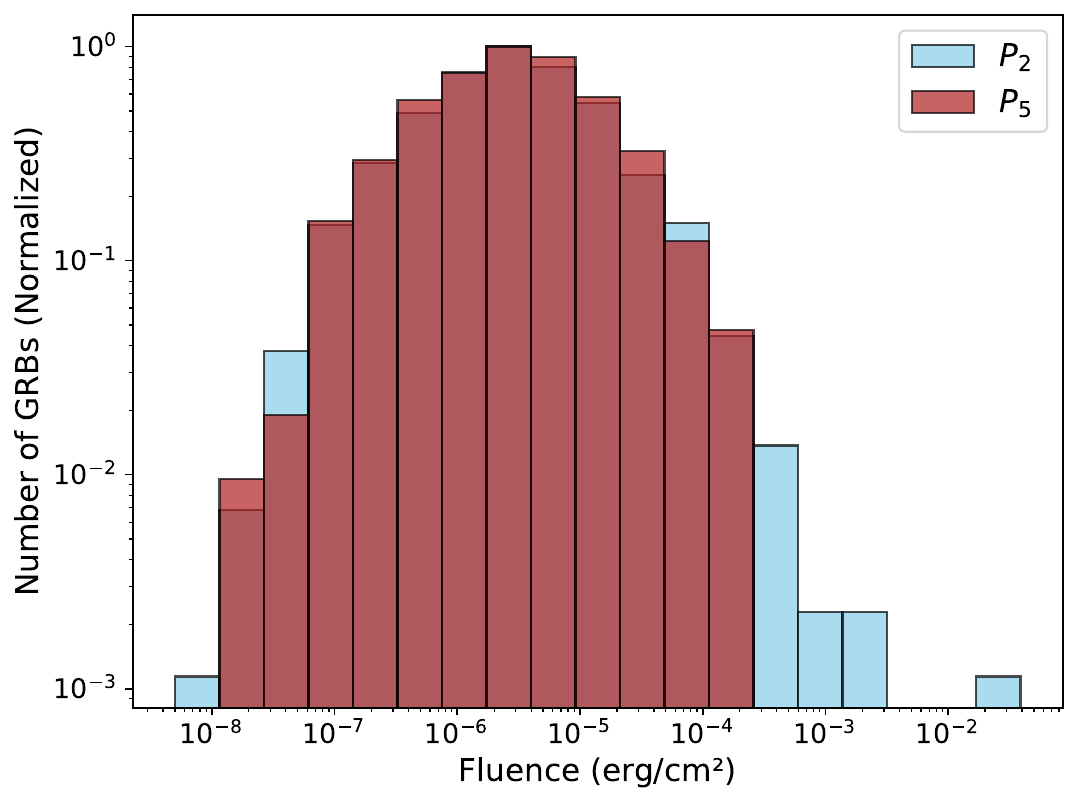}
    \caption{Comparison between the simulated GRB population and the observed catalogs. Left panel: normalized redshift distributions of the populations $P_{1}$ and $P_{5}$. The vertical dashed lines indicate the mean redshift of each sample, showing close agreement between the two distributions. Right panel: normalized fluence distributions of $P_{2}$ and $P_{5}$, illustrating that the mock sample reproduces the prompt-emission properties of GBM-detected GRBs.}
    \label{fig:GRB_catagorization}
\end{figure*}

In this section, we compare the properties of the simulated GRB population, $P_{5}$, with those of the observed GRB catalogs used in this work. The redshift and fluence distribution of $P_5$ are shown in Fig.~\ref{fig:detection_limit}.
Fig.~\ref{fig:GRB_catagorization} (left panel) shows the redshift distribution of the GRB population with measured redshift ($P_{1}$) and of the simulated GRB population ($P_{5}$). Both distributions are normalized to unity. The two populations show the same shape, with mean redshift of $\langle z \rangle = 1.99$ for $P_{1}$ and $\langle z \rangle = 1.91$ for $P_{5}$. The right panel of Fig.~\ref{fig:GRB_catagorization} compares the fluence distribution of the full \textit{Fermi}/GBM catalog ($P_{2}$) and $P_{5}$. Both distributions peak at $S_{\gamma} \sim 10^{-6}$–$10^{-5},\mathrm{erg,cm^{-2}}$ and show similar behavior over several orders of magnitude in fluence, with an exception for high fluences values. This can be explained by the fact that we are sampling few GRBs compared to the all GRBs detected by GBM, and very bright GRBs are typically less frequent.

\section{Independence between redshift and fluence}
In this section we quantify the statistical relationship between the redshift and the prompt fluence used to construct the mock GRB population $P_{5}$. As discussed in the main text, our procedure draws $z$ and $S_{\gamma}$ independently from $P_{1}$ and $P_{2}$, respectively, and then applies a detectability cut based on the $E_{\rm iso}^{\gamma}$--$z$ relation. Here we test whether this independence assumption is supported by the observed data.

\begin{figure}[h]
    \centering
    \includegraphics[width=\columnwidth]{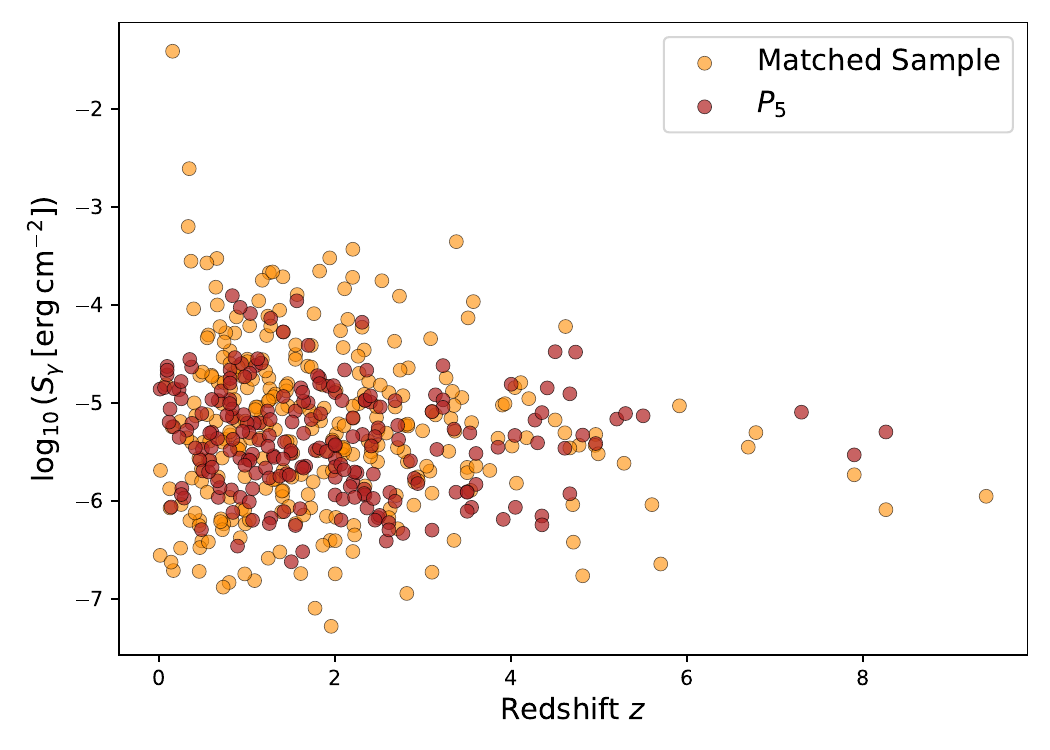}
    \caption{Fluence vs redshift for the Matched sample (orange) and the mock population $P_5$ (dark red). Both the distributions do not show strong signs of correlation between $z$ and $S_{\gamma}$.}
    \label{fig:z_fluence_scatter}
\end{figure}

We constructed a cross-matched sample by associating each GRB in the Greiner catalog with a measured redshift to its corresponding fluence (when possible) in the \textit{Fermi}/GBM catalog. This yields a sample of $N=297$ GRBs with both quantities measured for the same physical event, hereafter referred to as the matched sample. Fig.~\ref{fig:z_fluence_scatter} shows the resulting $(z, S_{\gamma})$ distribution for the matched sample, together with the corresponding distribution for the simulated mock population $P_{5}$, for visual comparison.

On the matched sample, we tested for a statistical dependence between $z$ and $\log(S_{\gamma})$ using three complementary tests: a Pearson linear correlation coefficient, a Spearman rank correlation coefficient,  and a two-dimensional Kolmogorov-Smirnov (KS) test \citep{Peacock1983, Fasano1987}. For the latter, we compared the observed joint distribution against a reference sample obtained by randomly permuting the fluence values relative to redshift, which enforces independence by construction while preserving both marginal distributions; the reported statistic is the median over 200 independent permutations. The results are summarized in Table~\ref{tab:independence_tests}.

None of the three tests reveal a statistically significant dependence between $z$ and $S_{\gamma}$ ($p>0.05$ in all cases), supporting the validity of treating the two quantities as independent when sampling the mock population.

\begin{table}
\caption{Independence tests between redshift and fluence on the Greiner-GBM matched sample ($N=297$).}
\label{tab:independence_tests}
\centering
\begin{tabular}{lcc}
\hline\hline
Test & Statistic & $p$-value \\
\hline
Pearson correlation & $r=-0.
098$ & 0.093 \\
Spearman rank correlation & $\rho=-0.039$ & 0.507 \\
2D Kolmogorov--Smirnov$^{\ddagger}$ & $D=0.049$ & 0.963 \\
\hline
\end{tabular}
\end{table}

\section{Observational Strategy}
\begin{figure}[h]
    \centering
    \includegraphics[width=\columnwidth]{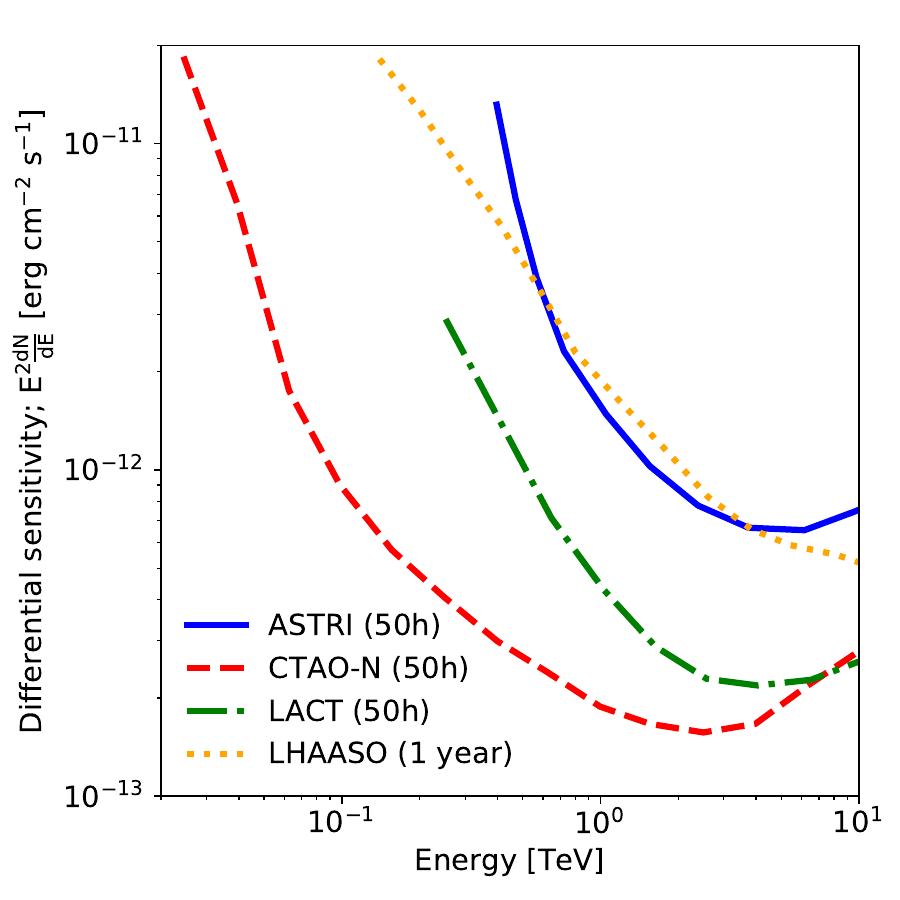}
    \caption{The sensitivity curves for ASTRI \citep{2024Univ...10..146S}, LACT \citep{Zhang:2025Ho}, and LHAASO. In addition, we also show the sensitivity of CTAO-N (obtained from \href{https://www.ctao.org/for-scientists/performance/}{CTAO performance documentation}). All the sensitivities are obtained under standard assumptions, such as, dark sky condition, low zenith angle {($20^\circ$)} and high atmospheric transmission.}
    \label{fig:Detector_sensitivity}
\end{figure}

\begin{figure}[h]
    \centering
    \includegraphics[width=\columnwidth]{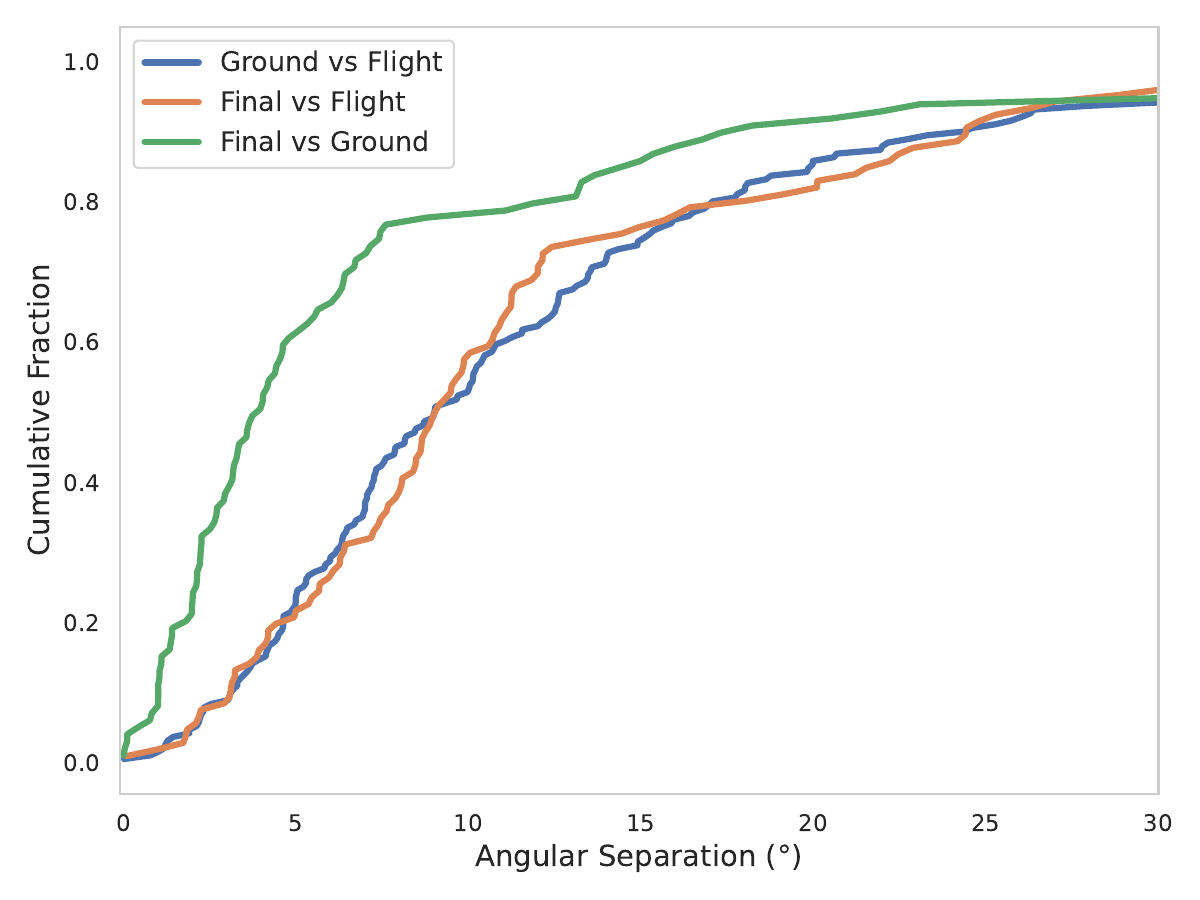}
    \caption{Cumulative distribution of the angular separation between different Flight, Ground and Final GCN localizations for GRBs detected by \textit{Fermi}/GBM. The curves show the angular separation between the onboard Flight localization and the Ground refined localization (blue), between the Flight and the Final (orange), and between Ground and Final (green). The distributions highlight the typical magnitude of position updates as localization information improves over time.}
    \label{fig:AngSep_GCN}
\end{figure}

\begin{figure}[h]
    \centering
    \includegraphics[width=\columnwidth]{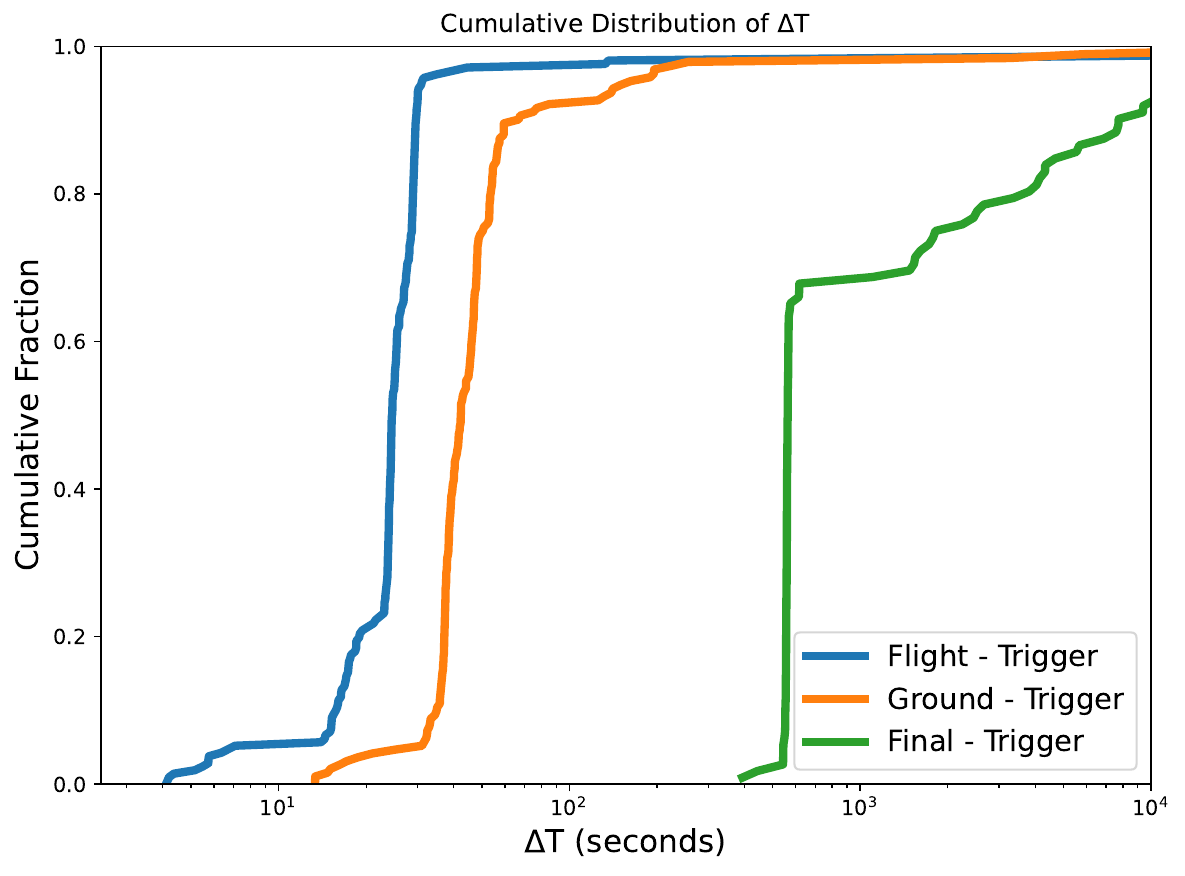}
    \includegraphics[width=\columnwidth]{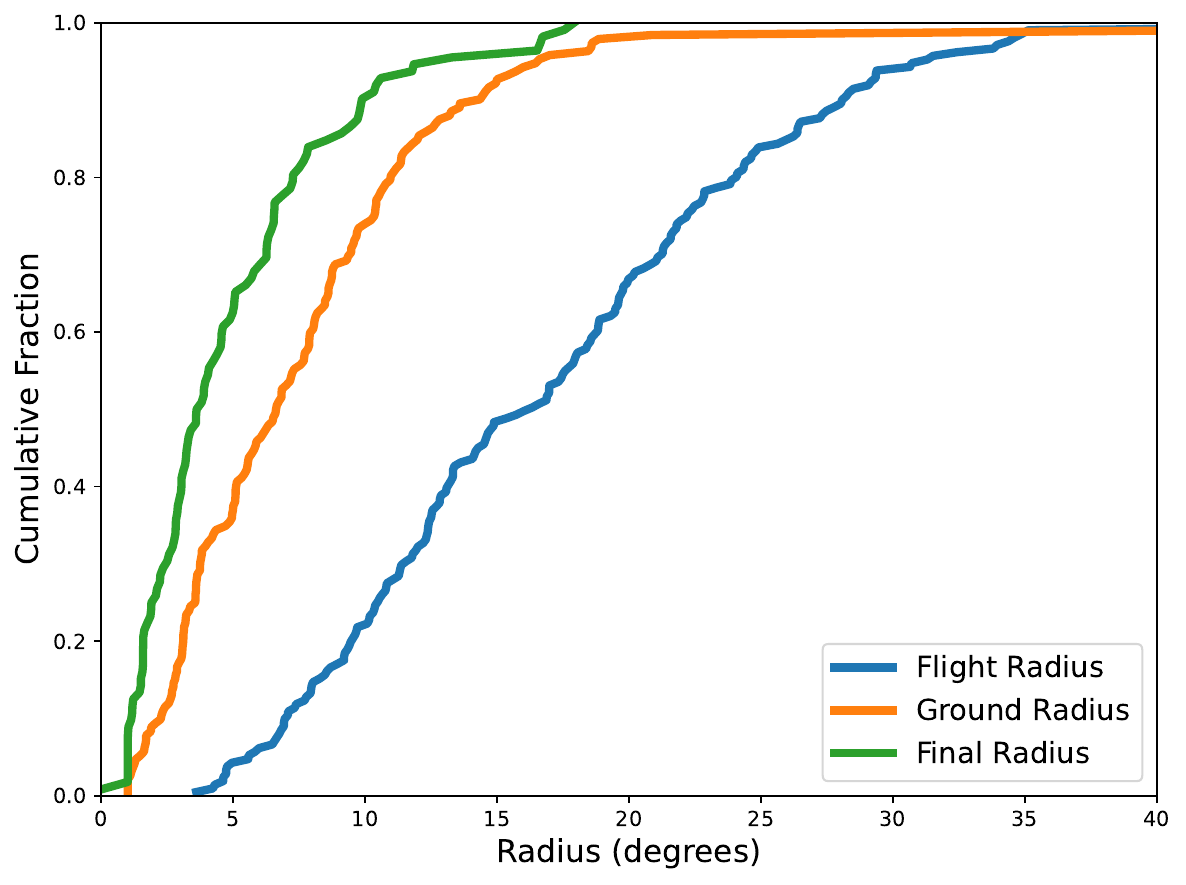}
    \caption{Cumulative distributions of the latency and localization accuracy of GCN notices for \textit{Fermi}/GBM detected GRBs. \textit{Top panel}: Distribution of the time delay ($\Delta T$) between the GRB trigger and the Flight notice, first Ground notice, and Final notice. \textit{Bottom panel}: Distribution of the corresponding localization error radii for the Flight, Ground, and Final notices. These distributions quantify the typical timescales and improvements in the localization relevant for planning rapid follow-up observations with ground based telescopes.}
    \label{fig:GCN_info_t_R}
\end{figure}

\begin{figure}
    \centering
    \includegraphics[width=\columnwidth]{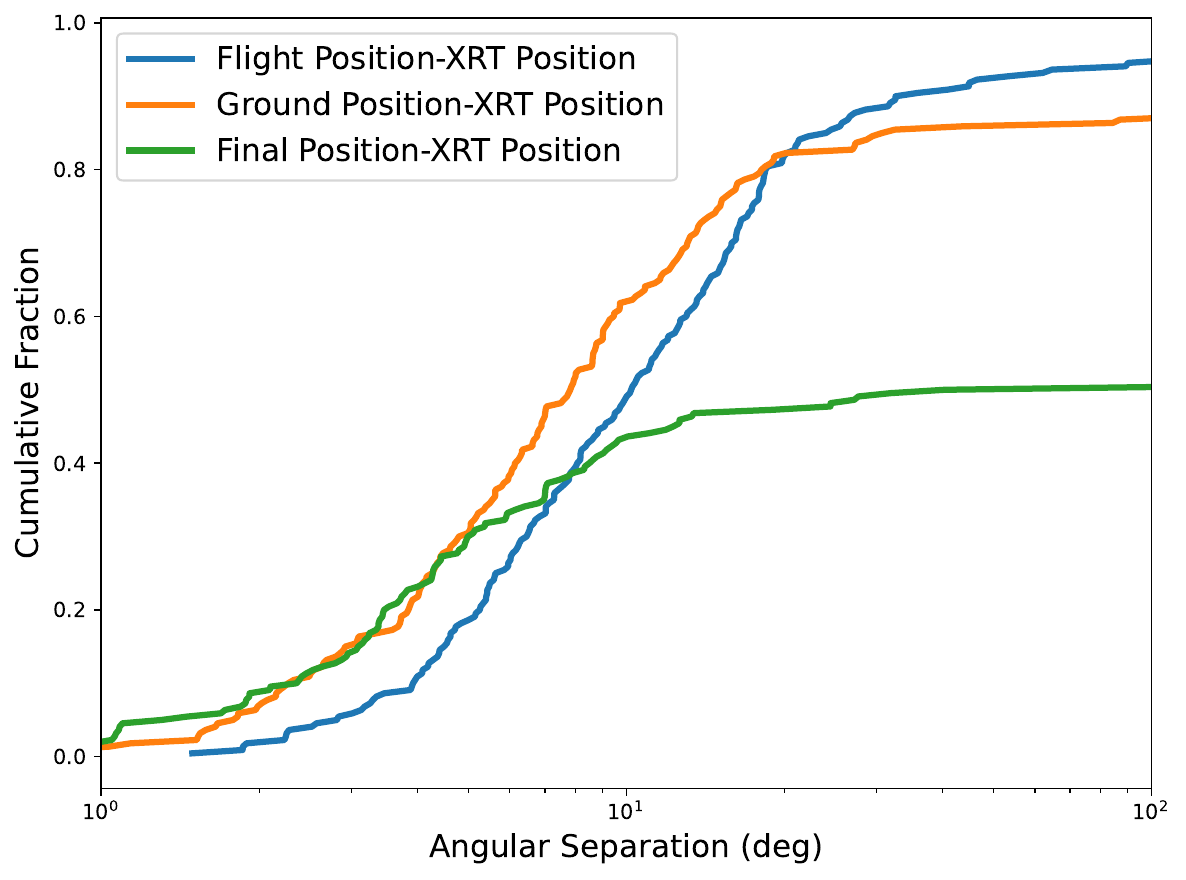}
    \caption{Cumulative distribution of the angular separation between the GRB positions reported in successive \textit{Fermi}/GBM GCN notices and the accurate sky position measured by \textit{Swift}/XRT. This figure shows the progressive improvement in localization accuracy as refined GCN notices become available and quantifies the typical angular offsets relevant for follow-up observations.}
    \label{fig:xrt-gbm separation}
\end{figure}

\subsection{very-high-energy gamma-ray instruments}
We computed the sensitivity of different detectors as presented in Tab.\ref{tab:Detectors}. The detector sensitivities are shown in Fig.~\ref{fig:Detector_sensitivity}.

\subsection{Trigger properties}\label{secA:trigger_properties}
A cumulative distribution of the quantities, namely $\rm t_{fl}, R_{fl}, t_{g}, R_{g}, t_{fi}, and~R_{fi}$ provides meaningful information. Figure \ref{fig:AngSep_GCN} presents the cumulative distribution of angular separation between the central point of the localization of flight-ground ($\rm \Delta R_{fl, g}$, blue), ground-final ($\rm \Delta R_{g,fi}$, green), and flight-final ($\rm \Delta R_{fl, fi}$, orange). 
This results in 68 percentile of the angular separation between the flight and ground being $\rm \Delta R_{fl, g}$ $\leq 13^{\circ}$, and between ground and final $\rm \Delta R_{g, fi}$ $\leq 6^{\circ}$.
Figure \ref{fig:GCN_info_t_R} shows the cumulative distribution of the latency of the arrival time of the notices (upper panel; latency of the flight notice: $\rm t_{fl}$, ground notice: $\rm t_{g}$, and final notice: t$_{fi}$), and the radius of the sky-localization  (lower panel; flight notice: $\rm R_{fl}$, ground notice: $\rm R_{g}$, and final notice: R$\rm_{fi}$). 
We extract the 68\% percentile of the above quantities which results in $\rm t^{68}_{fl}\leq30\,s, R^{68}_{fl}\leq 20^\circ, t^{68}_{g}\leq 50\,s, R^{68}_{g}\leq 10^{\circ}, t^{68}_{fi}\leq 600\,s,and~R^{68}_{fi}\leq 7^{\circ}$.

\begin{figure}[ht!]
    \centering
    \includegraphics[width=\columnwidth]{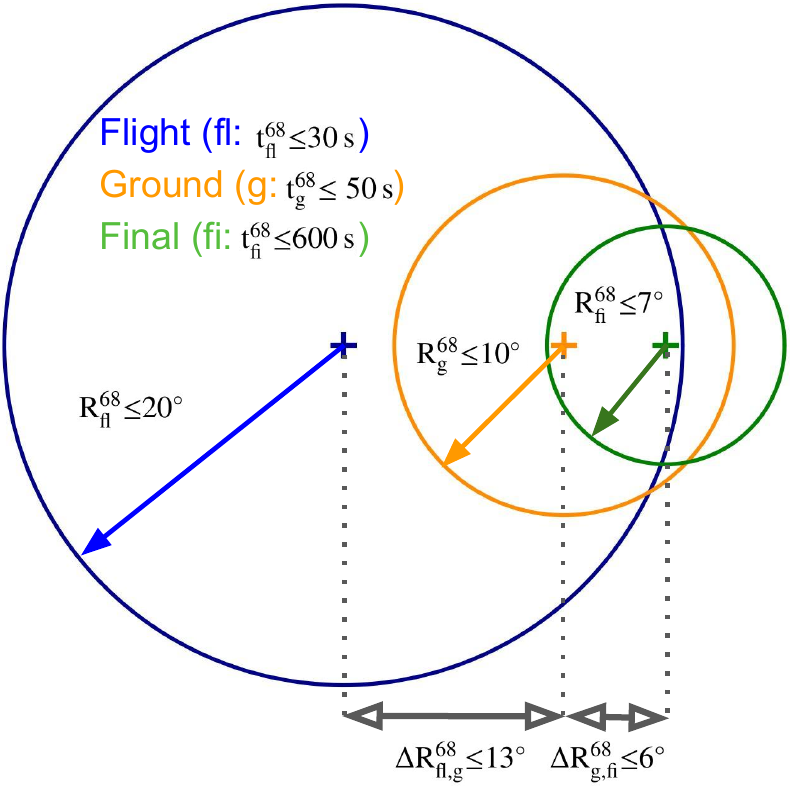}
    \caption{Comparison of the information distributed by flight, ground and final notices. The a significantly improved uncertainty of the sky-area is comunicated through the ground alerts which are distributed typically about a minute after the GRB trigger.}
    \label{fig:GCN_info1}
\end{figure}

The information discussed previously can be summarized in Fig.~\ref{fig:GCN_info1}, where the blue circle is a representative of the sky-localization during the flight-trigger received with a typical time of 30\,s (or earlier) with a localization radius of about 20$^{\circ}$ (or smaller). An additional follow-up notice (ground-notice, indicated by the orange circle) is expected at about 50\,s (or earlier) with an angular separation of about 13$^{\circ}$ (or smaller) with a localization radius of 10$^{\circ}$ (or smaller). The following notice, the final notice (indicated by the green circle), arriving with a latency of about 600 \,s provides a localization with a radius of about 7$^{\circ}$ (or smaller) with an angular separation of about 6$^{\circ}$ (or less).

\section{Detection horizon}

\begin{figure}[h]
    \centering
    \includegraphics[width=\columnwidth]{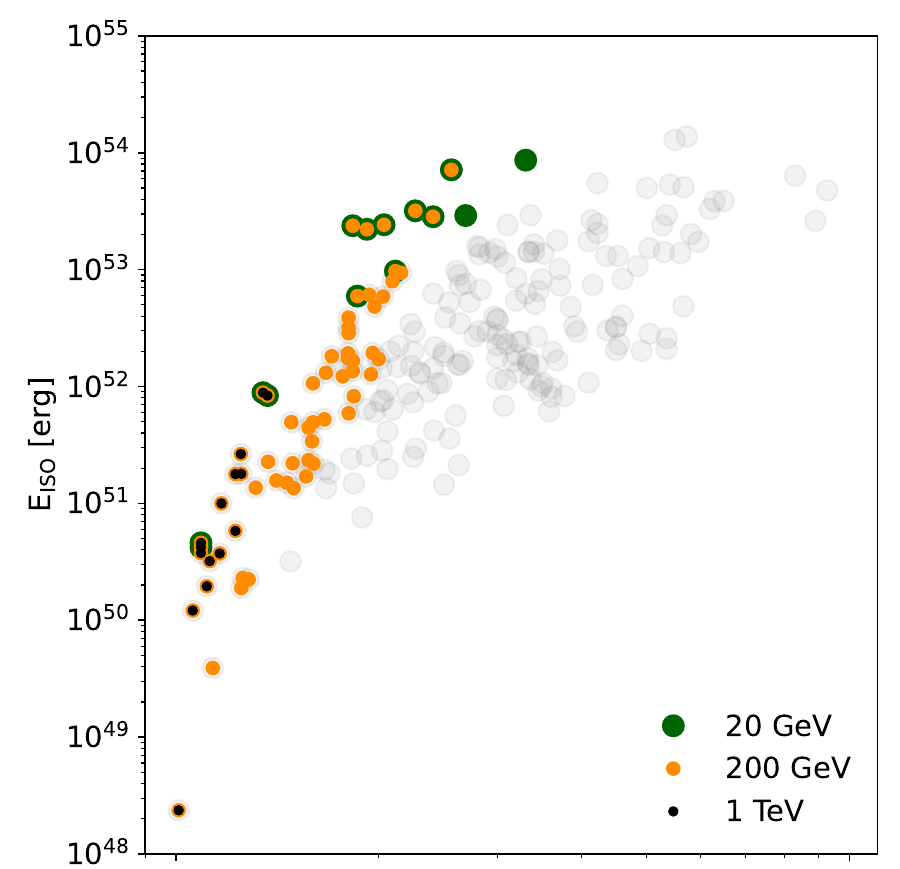}
     \includegraphics[width=\columnwidth]{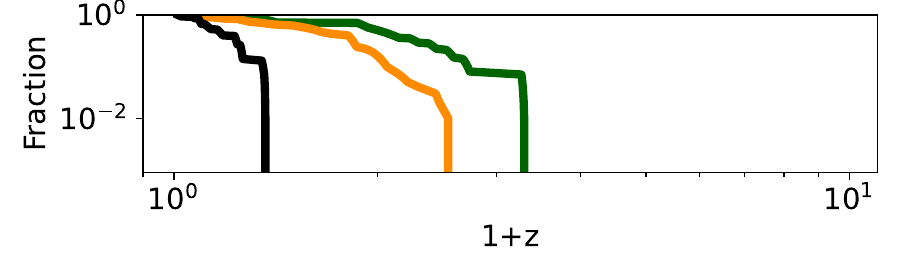}
    \caption{Detection horizon for CTAO-N telescope for different energy thresholds of 0.02 TeV (green), 0.2 TeV (orange) and 1 TeV (black). They gray points are representative of the population of the total synthetic population. For the horizon calculations, we assumed no duty cycles of the detectors. We further assumed that all the GRBs are followed up from 100\,s after the trigger.}
    \label{fig:Det_horizon}
\end{figure}

We estimated the detection horizon for a synthetic population of GRBs with assigned redshifts and E$_{\rm ISO}$. We fixed the exposure to 5~minutes. Among the detectors listed in Tab.~\ref{tab:Detectors} (see also Sect.~\ref{sec:Detectors}), the sensitivity of CTAO-N is superior throughout the entire energy band between 20\,GeV--10\,TeV. Thus, we take CTAO-N as an example for estimating the detection horizon. We further assume that the telescope points to the location of the GRB starting from 100\,s which helps detecting even the fainter GRBs (with lower E$\rm _{ISO}$. We do not consider the duty cycle of the instrument for this computation. The detectability is computed based on the TeV light curves discussed in Sect.\ref{sec:LC_VHE_model} and a detection is claimed is the flux is above $\rm F^{5\,min}_{lim}(E)$ (see Eq.~\ref{eq:Flim}), where  three values of the energy, {20 GeV, 200 GeV} and 1 TeV, are chosen (see right panel of Fig.~\ref{fig:VHE_LCs}). 
As can been seen in Fig.~\ref{fig:Det_horizon}, the energy at which the detection is made (can be considered as the energy threshold) impacts the detection horizon. Most importantly, although the sensitivity at lower energies (E$\sim$20 GeV) is an order of magnitude worse than that of medium energy band (E$\sim$200 GeV), the EBL attenuation is subdominant. Hence a larger horizon of about redshift 4.0 can potentially be achieved. However, the increasing energy threshold restricts the horizon to $\sim$1.5 to $\sim$0.5, for 200 GeV, 1 TeV respectively.

\begin{figure}[h]
    \centering
    \includegraphics[width=\columnwidth]{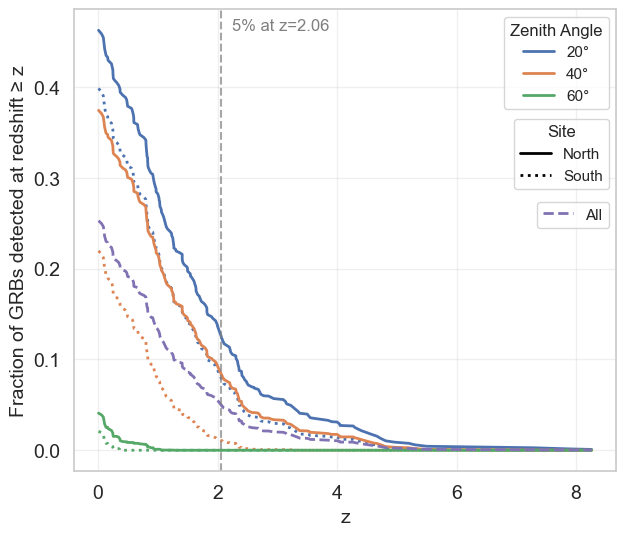}
    \caption{Fraction of detectable events with CTAO-N for increasing redshift, marginalized over latencies up to $4$~hr. Once the threshold of $z\sim2$ if reached, fewer than 5~\% of events are detectable. At higher zenith angles, few events are detectable at the $5\sigma$ level.}
    \label{fig:Det_horizon_CTAO}
\end{figure}

When considering the full energy range and time-dependent light curves of the events as simulated in Sect.~\ref{sec:ctao-sensipy}, we see a compatible horizon for follow-up campaigns up to 4~hrs long. As shown in Fig~\ref{fig:Det_horizon_CTAO}, the fraction of detectable events drops rapidly with increasing redshift for all observatory configurations. Above $z\sim2$, only 5~\% of the events can still be detected in all sites and zenith ranges. In particular, fewer than 5~\% of events can be detected at a zenith $\ge60^{\circ}$, which can be used to inform the followup criteria for CTAO.

\end{document}